\documentclass[aps,prc,twocolumn,superscriptaddress,showpacs]{revtex4} %not used

\usepackage{graphicx} % used
\usepackage{draftcopy} %not used
\usepackage{afterpage} %used

\usepackage{lineno} % used
\newcommand{\agev}    {\mbox{$A$~GeV}}               %PRL notation
\newcommand{\gevc}    {\mbox{GeV$/c$}}

\newcommand{\mevc}    {\mbox{MeV$/c$}}

\newcommand{\dedx}    {\ensuremath{\textrm{d}E/\textrm{d}x}}
\newcommand{\dndy}    {\ensuremath{\textrm{d}N/\textrm{d}y}}

\newcommand{\der}     {\ensuremath{\textrm{d}}}

\newcommand{\tpm}     {\ensuremath{\! \pm \!}}

\begin{document}
%\Blindtext
%\preprint{}

%==========================================================================

\title{Production of deuterium, tritium, and $^3$He in central Pb+Pb collisions
at 20$A$, 30$A$, 40$A$, 80$A$, and 158\agev~at the CERN SPS}


%\author{}
%\thanks{}
%\affiliation{}
%===========================================================================%
% List of institutions
%
\affiliation{NIKHEF,
             Amsterdam, Netherlands.}
\affiliation{Department of Physics, University of Athens, 
             Athens, Greece.}
\affiliation{E\"otv\"os Lor\'ant University,
	    Budapest, Hungary.}
\affiliation{KFKI Research Institute for Particle and Nuclear Physics,
             Budapest, Hungary.} 
\affiliation{MIT,
             Cambridge, USA.}
\affiliation{H.~Niewodnicza\'nski Institute of Nuclear Physics, Polish Academy of Sciences,
	    Cracow, Poland.}
\affiliation{Gesellschaft f\"{u}r Schwerionenforschung (GSI),
             Darmstadt, Germany.}
\affiliation{Joint Institute for Nuclear Research,
             Dubna, Russia.}
\affiliation{Fachbereich Physik der Universit\"{a}t,
             Frankfurt, Germany.}
\affiliation{CERN,
             Geneva, Switzerland.}
\affiliation{Institute of Physics, Jan Kochanowski University,
	     Kielce, Poland.}
\affiliation{Fachbereich Physik der Universit\"{a}t, 
             Marburg, Germany.}
\affiliation{Max-Planck-Institut f\"{u}r Physik,
             Munich, Germany.}
\affiliation{Inst. of Particle and Nuclear Physics, Charles Univ.,
	    Prague, Czech Republic.}
\affiliation{Nuclear Physics Laboratory, University of Washington,
             Seattle, WA, USA.}
\affiliation{Atomic Physics Department, Sofia University St.~Kliment Ohridski,
             Sofia, Bulgaria.}
\affiliation{Institute for Nuclear Research and Nuclear Energy, BAS,
             Sofia, Bulgaria.}
\affiliation{Department of Chemistry, Stony Brook Univ. (SUNYSB),
             Stony Brook, USA.}
\affiliation{Institute for Nuclear Studies,
             Warsaw, Poland.}
\affiliation{Institute for Experimental Physics, University of Warsaw,
             Warsaw, Poland.}
\affiliation{Faculty of Physics, Warsaw University of Technology,
             Warsaw, Poland.}
\affiliation{Rudjer Boskovic Institute,
             Zagreb, Croatia.}
%===========================================================================%
% Author list
%
\author{T.~Anticic}
\affiliation{Rudjer Boskovic Institute,
             Zagreb, Croatia.}
\author{B.~Baatar}
\affiliation{Joint Institute for Nuclear Research,
             Dubna, Russia.}
\author{J.~Bartke}
\altaffiliation{deceased}
\affiliation{H.~Niewodnicza\'nski Institute of Nuclear Physics, Polish Academy of Sciences,
	    Cracow, Poland.}
\author{H.~Beck}
\affiliation{Fachbereich Physik der Universit\"{a}t,
             Frankfurt, Germany.}
\author{L.~Betev}
\affiliation{CERN,
             Geneva, Switzerland.}
\author{H.~Bia{\l}\-kowska}
\affiliation{Institute for Nuclear Studies,
             Warsaw, Poland.}
\author{C.~Blume}
\affiliation{Fachbereich Physik der Universit\"{a}t,
             Frankfurt, Germany.}
\author{B.~Boimska}
\affiliation{Institute for Nuclear Studies,
             Warsaw, Poland.}
\author{J.~Book}
\affiliation{Fachbereich Physik der Universit\"{a}t,
             Frankfurt, Germany.}
\author{M.~Botje}
\affiliation{NIKHEF,
             Amsterdam, Netherlands.}
\author{P.~Bun\v{c}i\'{c}}
\affiliation{CERN, 
             Geneva, Switzerland.}
\author{P.~Christakoglou}
\affiliation{NIKHEF,
             Amsterdam, Netherlands.}
\author{P.~Chung}
\affiliation{Department of Chemistry, Stony Brook Univ. (SUNYSB),
             Stony Brook, USA.}
\author{O.~Chvala}
\affiliation{Inst. of Particle and Nuclear Physics, Charles Univ.,
	    Prague, Czech Republic.}
\author{J.G.~Cramer}
\affiliation{Nuclear Physics Laboratory, University of Washington,
             Seattle, WA, USA.}
\author{V.~Eckardt}
\affiliation{Max-Planck-Institut f\"{u}r Physik,
             Munich, Germany.}
\author{Z.~Fodor}
\affiliation{KFKI Research Institute for Particle and Nuclear Physics,
             Budapest, Hungary.}
\author{P.~Foka}
\affiliation{Gesellschaft f\"{u}r Schwerionenforschung (GSI),
             Darmstadt, Germany.}
\author{V.~Friese}
\affiliation{Gesellschaft f\"{u}r Schwerionenforschung (GSI),
             Darmstadt, Germany.}
\author{M.~Ga\'zdzicki}
\affiliation{Fachbereich Physik der Universit\"{a}t,
             Frankfurt, Germany.}
\affiliation{Institute of Physics, Jan Kochanowski University,
	     Kielce, Poland.}
\author{K.~Grebieszkow}
\affiliation{Faculty of Physics, Warsaw University of Technology, 
             Warsaw, Poland.}
\author{C.~H\"{o}hne}
\affiliation{Gesellschaft f\"{u}r Schwerionenforschung (GSI),
             Darmstadt, Germany.} 
\author{K.~Kadija}
\affiliation{Rudjer Boskovic Institute, 
             Zagreb, Croatia.}
\author{A.~Karev}
\affiliation{CERN,
             Geneva, Switzerland.}
\author{V.I.~Kolesnikov}
\affiliation{Joint Institute for Nuclear Research,
             Dubna, Russia.}
\author{M.~Kowalski}
\affiliation{H.~Niewodnicza\'nski Institute of Nuclear Physics, Polish Academy of Sciences,
	    Cracow, Poland.}
\author{D.~Kresan}
\affiliation{Gesellschaft f\"{u}r Schwerionenforschung (GSI),
             Darmstadt, Germany.}
\author{A.~Laszlo}
\affiliation{KFKI Research Institute for Particle and Nuclear Physics,
             Budapest, Hungary.} 
\author{R.~Lacey}
\affiliation{Department of Chemistry, Stony Brook Univ. (SUNYSB),
             Stony Brook, USA.}
\author{M.~van~Leeuwen}
\affiliation{NIKHEF,
             Amsterdam, Netherlands.}
\author{M.~Ma\'{c}kowiak-Paw{\l}owska}
\affiliation{Faculty of Physics, Warsaw University of Technology, 
             Warsaw, Poland.}
\author{M.~Makariev}
\affiliation{Institute for Nuclear Research and Nuclear Energy, BAS,
             Sofia, Bulgaria.}
\author{A.I.~Malakhov}
\affiliation{Joint Institute for Nuclear Research,
             Dubna, Russia.}
\author{G.L.~Melkumov}
\affiliation{Joint Institute for Nuclear Research,
             Dubna, Russia.}
\author{M.~Mitrovski}
\affiliation{Fachbereich Physik der Universit\"{a}t, 
             Frankfurt, Germany.}
\author{St.~Mr\'owczy\'nski}
\affiliation{Institute of Physics, Jan Kochanowski University,
	     Kielce, Poland.}
\author{G.~P\'{a}lla}
\affiliation{KFKI Research Institute for Particle and Nuclear Physics,
             Budapest, Hungary.} 
\author{A.D.~Panagiotou}
\affiliation{Department of Physics, University of Athens,
             Athens, Greece.}
\author{D.~Prindle}
\affiliation{Nuclear Physics Laboratory, University of Washington,
             Seattle, WA, USA.}
\author{F.~P\"{u}hlhofer}
\affiliation{Fachbereich Physik der Universit\"{a}t,
             Marburg, Germany.}
\author{R.~Renfordt}
\affiliation{Fachbereich Physik der Universit\"{a}t, 
             Frankfurt, Germany.}
\author{C.~Roland}
\affiliation{MIT,
             Cambridge, USA.}
\author{G.~Roland}
\affiliation{MIT, 
             Cambridge, USA.}
\author{A.~Rustamov}
\affiliation{Fachbereich Physik der Universit\"{a}t,
             Frankfurt, Germany.}
\author{M.~Rybczy\'nski}
\affiliation{Institute of Physics, Jan Kochanowski University,
	     Kielce, Poland.}
\author{A.~Rybicki}
\affiliation{H.~Niewodnicza\'nski Institute of Nuclear Physics, Polish Academy of Sciences,
	    Cracow, Poland.}
\author{A.~Sandoval}
\affiliation{Gesellschaft f\"{u}r Schwerionenforschung (GSI),
             Darmstadt, Germany.} 
\author{N.~Schmitz}
\affiliation{Max-Planck-Institut f\"{u}r Physik, 
             Munich, Germany.}
\author{T.~Schuster}
\affiliation{Fachbereich Physik der Universit\"{a}t, 
             Frankfurt, Germany.}
\author{P.~Seyboth}
\affiliation{Max-Planck-Institut f\"{u}r Physik,
             Munich, Germany.}
\author{F.~Sikl\'{e}r}
\affiliation{KFKI Research Institute for Particle and Nuclear Physics,
             Budapest, Hungary.} 
\author{E.~Skrzypczak}
\affiliation{Institute for Experimental Physics, University of Warsaw,
             Warsaw, Poland.}
\author{M.~Slodkowski}
\affiliation{Faculty of Physics, Warsaw University of Technology,
             Warsaw, Poland.}
\author{G.~Stefanek}
\affiliation{Institute of Physics, Jan Kochanowski University,
	     Kielce, Poland.}
\author{R.~Stock}
\affiliation{Fachbereich Physik der Universit\"{a}t,
             Frankfurt, Germany.}
\author{H.~Str\"{o}bele}
\affiliation{Fachbereich Physik der Universit\"{a}t,
             Frankfurt, Germany.}
\author{T.~Susa}
\affiliation{Rudjer Boskovic Institute,
             Zagreb, Croatia.}
\author{M.~Szuba}
\affiliation{Faculty of Physics, Warsaw University of Technology,
             Warsaw, Poland.}
\author{D.~Varga}
\affiliation{E\"otv\"os Lor\'ant University,
	    Budapest, Hungary.}
\author{M.~Vassiliou}
\affiliation{Department of Physics, University of Athens, 
             Athens, Greece.}
\author{G.I.~Veres}
\affiliation{KFKI Research Institute for Particle and Nuclear Physics,
             Budapest, Hungary.}
\author{G.~Vesztergombi}
\affiliation{KFKI Research Institute for Particle and Nuclear Physics,
             Budapest, Hungary.}
\author{D.~Vrani\'{c}}
\affiliation{Gesellschaft f\"{u}r Schwerionenforschung (GSI),
             Darmstadt, Germany.}
\author{Z.~W{\l}odarczyk}
\affiliation{Institute of Physics, Jan Kochanowski University,
	     Kielce, Poland.}
\author{A.~Wojtaszek-Szwarc}
\affiliation{Institute of Physics, Jan Kochanowski University,
	     Kielce, Poland.}
%===========================================================================
\collaboration{The NA49 Collaboration}
\noaffiliation

%\date{\today \\  {\bf ..DRAFT 0.0}} %used

%===========================================================================

\begin{abstract}
Production of $d$, $t$,  and $^3$He nuclei in central
Pb+Pb interactions was studied at five collision energies ($\sqrt{s_{NN}}=$~6.3, 7.6,
8.8, 12.3, and 17.3~GeV) with the NA49 detector at the CERN SPS. 
Transverse momentum spectra, rapidity distributions, and particle ratios were measured. 
Yields are compared to predictions of statistical models.
Phase-space distributions of light nuclei are discussed and compared to those
of protons in the context of a coalescence approach. The coalescence parameters $B_2$ and $B_3$,
as well as coalescence radii for $d$ and $^3$He were determined as a function of transverse mass
at all energies.
\end{abstract}

%=========% insert suggested PACS numbers in braces on next line
\pacs{25.75.Dw}
%\keywords{}
\maketitle

\section{Introduction}
The main goal of the heavy-ion program at the CERN SPS is the experimental
investigation of the properties of nuclear matter under extreme conditions.
In a head-on inelastic collision of lead nuclei, accelerated to an energy
of several tens of GeV per nucleon, a hot and dense fireball
of an extraordinary outward pressure gradient is formed.
After explosion-like decompression, the fireball expands
well beyond the volume defined by the geometric overlap region of the colliding nuclei
resulting in a multiparticle system with strong collective behavior.
Experimentally the overall dynamical evolution of the reaction
can be probed by measuring particle composition, longitudinal and
transverse momentum distributions of different particle species, as well
as multi-particle correlations.
  
The study of light nuclei production is of importance for
several reasons. First of all, the mechanism of cluster formation in the
interior of the fireball of a heavy-ion collision is not well understood and requires
further quantitative investigations. It is likely that a significant fraction
of few-nucleon bound states registered near mid-rapidity are produced in a late stage
of the reaction when the hadronic matter becomes diluted and most of the newly
formed hydrogen and helium isotopes decouple from the source having no
subsequent rescatterings. So, light nuclei may serve as probes of the
fireball dynamics at the time of the freezeout. 

In the simplest coalescence
model~\cite{coal1,coal2,mrozhin} the yields of light nuclei are well explained
as being determined solely by the distributions of their constituents (protons and
neutrons) and an empirical coalescence parameter ($B_A$) related to the size $A$ of the cluster. 
Such an approach works very well in proton-induced reactions and in nuclear
interactions at low energies where collective flow effects are smaller. However,
for expanding nuclear matter, the simple coalescence model needs to be modified.
In relativistic heavy-ion collisions the production of nucleon composites depends on the
reaction "homogeneity volume"~\cite{sinyukov}, whose characteristics are
likely to be sensitive not only to the nucleon phase space distributions at
freezeout,  but also to the strength of momentum-space correlations induced
by collective flow~\cite{heinz}. To get insight into the structure of the
source and the characteristics of its density and flow velocity profile, the parameters of the rapidity and
transverse momentum distributions of clusters of different sizes need to be obtained over
a large phase space region.
        
Another commonly employed approach to describe particle yields is based on statistical
and thermal hadro-chemical equilibrium models of particle production~\cite{capusta,mekyan,shm,therm1}.
In a conventional thermal model, particle multiplicities are predicted
dependent on the bulk thermal parameters of the system - the chemical freezeout temperature
$T$, baryochemical potential $\mu_B$ and volume $V$. Though recent versions of statistical
thermal models are capable to describe hadron abundances  from heavy-ion
reactions in the range of collision energies from about 1 to several
$10^3$~GeV per nucleon~\cite{therm2,therm3,therm4}, the question was raised~\cite{therm4,entr1,entr2}
whether this approach is justified when applied to the production of nucleon clusters.
The present paper does not discuss this issue, but follows most previous publications on light
nuclei production in applying the statistical model to the hadron composition
at freezeout and assuming that the yields remain unchanged during the further 
evolution of the fireball. Since thermal models basically consider
particle yields integrated over the full phase space the lack of results on
total multiplicities of light nuclei has up to now prevented a straightforward and
quantitative test of the applicability of the statistical model approach to 
light nuclei production in the GeV collision energy range. 
Thanks to the large acceptance of the NA49 experiment,
$^3$He and $d$ production can be measured and analyzed in 
a significant part of the final-state phase space
allowing an extrapolation of the yields to full (4$\pi$) phase space.
This opens the possibility to examine the statistical approach by these
little explored experimental probes.

The study of the production of light nuclei with different proton to neutron
ratios in heavy-ion collisions can probe the behavior of the asymmetric
dense nuclear matter Equation-of-State (EOS) for a range of densities,
temperatures and proton fractions.
The  dense nuclear matter EOS is of fundamental importance
for both nuclear physics and astrophysics. 
Several experimental observables which potentially reveal information about
the density dependence of the symmetry energy associated with the $n-p$ asymmetry
("symmetry energy" term in the nuclear EOS) have been proposed at low and 
intermediate collision energies: multifragmentation~\cite{mult1,mult2,mult3},
nucleon directed and elliptic flow~\cite{nucl_flow}, the charged pion ratio
$\pi^-/\pi^+$~\cite{pions}, and the isobaric yield ratios of light
clusters~\cite{isobar}. Unfortunately, none of these probes is uniquely sensitive
to the symmetry energy at all nuclear densities. For example, calculations
within an isospin-dependent transport model demonstrated that the
$n/p$ ratio is most sensitive to the symmetry energy at subnormal
densities (low collision energies), while with increasing collision energy
(at supra-normal densities) the sensitivity of the $\pi^-/\pi^+$ ratio to the
density dependence of the symmetry energy was found to be stronger~\cite{esym_pions}. 
Thus, in order to map out the entire density dependence of the 
symmetry energy, a combination of several complementary observables 
in a broad range of collision energies is necessary. 
NA49 is not capable of detecting and identifying neutrons.
Under certain circumstances, however, one might
expect that the yield ratio of tritons to $^3$He is a measure of the $n/p$ ratio in the fireball,
because the freezeout nucleon isospin asymmetry leads to a sizable difference
in the production rates for light nuclei of different nucleon composition. 
This paper presents the results of a study comparing the production rates of
$A=3$ nuclei in the SPS energy range.

It should be noted that when production of a composite of mass number
$A$ is analyzed in the framework of a coalescence approach, the most common
assumption is that the yield of neutrons is equal to that of protons.
This is only partially true, because in relativistic heavy-ion collisions
relative abundances of nucleon species are expected to change considerably during
the dynamical evolution of the reaction  due to multiple rescattering
effects in dense hadronic matter. Indeed, at AGS energies a mid-rapidity
$n/p$ ratio  $R_{np}=1.19\pm0.08$ was obtained for central Au+Pb
collisions at 11.5\agev~\cite{neut_e864}, which differs from both
the $n/p$-ratio in the incident nuclei prior to the interaction
($\approx 1.5$) and $R_{np}=1$ used in coalescence studies. The
latter assumption introduces an extra systematic error into  the results for
the coalescence parameters $B_A$ which scales
as $R_{np}^{A-1}$. There is no experimental data on $R_{np}$ for heavy-ion
collisions above AGS energies since usually experiments have not been capable 
of detecting neutrons far from beam rapidity. Thus, new
experimental data on the energy dependence of the triton to $^3$He
asymmetry in the participant region of a central Pb+Pb collision can shed
some light on the degree of chemical equilibration attained at SPS energies.

Up to now, light (anti)nuclei production has been studied extensively only
in the energy range below $\sqrt{s_{NN}}\approx$~20~GeV
at the AGS~\cite{clust_e802,clust_e864,clust_e877,clust_e896} and
SPS~\cite{deu_na52,deu_na44,deu1_na49,dbar_na49}.
In Ref.~\cite{deu2_na49} the NA49 experiment reported mid-rapidity spectra of
deuterons from central Pb+Pb collisions at $\sqrt{s_{NN}}=8.8$, 12.3, and
17.3~GeV. This paper presents results for a wider range of
cluster species, which includes also tritons and $^3$He nuclei, measured in
central Pb+Pb interactions at center of mass energies from 6 to 17~GeV.   

The paper is structured as follows. Section II describes the NA49 experiment
and the studied data sets. Section III outlines the details
of the analysis procedure for light nuclei. The main results
of the paper are presented and discussed in Section IV. Section V concludes the paper
with a summary of the results.

\section{Experimental setup}

The NA49 detector is a large acceptance magnetic spectrometer for
the study of hadron production in heavy-ion collisions at the CERN SPS.
The detector components are described briefly below and a complete
description is given in Ref.~\cite{na49_setup}.
The tracking system consists of four Time-Projection Chambers (TPCs).
Two Vertex TPCs (VTPC) are placed inside two super-conducting dipole
magnets and provide momentum analysis.
Downstream of the magnets Main TPCs (MTPC) are positioned on each side
of the beam trajectory. These record the tracks
of charged particles providing up to 90 measurements of the position and
specific energy loss \dedx~of charged particles. A resolution of
$\sigma_{\dedx}\approx4\%$ is achieved for the MTPCs allowing identification
of charged particles in the relativistic rise region by correlating their
\dedx~and momentum.  Time-Of-Flight (TOF) detectors, composed each of 891 fast plastic
scintillator tiles, are placed behind each MTPC.
The TOF walls have a timing resolution of 60~ps and are essential for the
identification of deuterons and tritons up to momenta of 12~\gevc.
A zero-degree calorimeter (VCAL) located further
downstream is employed to trigger on collision centrality. 
The trajectory of incident beam ions is measured with three proportional
counters (BPD1, 2, 3). A set of scintillation and Cherenkov counters
positioned upstream of the target is used for beam definition and provides
the start of the timing of the experiment.
\begin{table}[tb]%[H] add [H] placement to break table across pages
\begin{center}
\caption{\label{Table1} Summary of the data sets used in the analysis.}
\begin{ruledtabular}
\begin{tabular}{ccccc}
E$_{beam}$ (\agev) & \hspace{1mm}$\sqrt{s_{NN}}$~(GeV)
 & \hspace{1mm}    centrality & $\langle N_W \rangle$
        & \hspace{1mm}$N_{events}$\hspace{1mm} \\
\hline
20 & 6.3 & 0-7\% & 349 & 350,000\\
30 & 7.6 & 0-7\% & 349 & 400,000\\
40 & 8.8 & 0-7\% & 349 & 700,000\\
80 & 12.3 & 0-7\% & 349 & 250,000\\
158 & 17.3 & 0-12\% & 335 &1200,000\\
\end{tabular}
\end{ruledtabular}
\end{center}
\end{table}

\section{Data analysis}

\subsection{Data sets}
The data used in this analysis were collected in years 1996-2002.
The experiment utilized a $^{208}$Pb beam at energies of 20$A$, 30$A$, 40$A$, 80$A$
and 158\agev~ impinging on a lead target
of 224 mg/cm$^2$ thickness corresponding to a 1\% interaction probability.
The interaction trigger selected the 12\% most central collisions
at 158\agev, at other beam energies the data were recorded with
a 7\% central trigger.
In order to obtain similar acceptance at all beam energies  
the strength of the magnetic field in the VTPCs
was changed in proportion to the beam energy. 
An overview of the data sets used in this analysis including collision energy,
centrality, and total number of events is given in Table~\ref{Table1}. 
Data at 40$A$ and 158\agev~ were recorded for two opposite polarities
of the magnetic field with approximately equal number of events for
each setting.

\subsection{Time-of-flight reconstruction}
As described in detail in Ref.~\cite{dbar_na49}, straight line (MTPC)
segments of reconstructed tracks were extrapolated to the TOF walls
and matched with TOF hits. Corrections for the position of the hit inside
the scintillator (tile) and the amplitude-dependent time-walk effect in the
discriminator were applied tile-wise. 
Values of mass-squared were then calculated from the
reconstructed  momentum $p$, the flight path to the TOF detector $l$ and the
measured time-of-flight $t$ as
\begin{equation}
  \label{eq:mass2}
  m^2=\frac{p^2}{c^2} \left( \frac{c^2t^2}{l^2}-1 \right)
\end{equation}
where $c$ denotes the speed of light.

\subsection{Event and track selection}

This section describes the cuts applied to select events and tracks for further analysis.
In order to reduce the background from non-target interactions,
only events for which the reconstructed primary vertex coordinate along the beam 
axis is within 1~cm from the nominal target position were retained. 
The fraction of events remaining after application of this cut
varied slightly with bombarding energy and was about 99\%.

To ensure optimal momentum resolution, tracks had to be reconstructed
in a VTPC and a MTPC and were required to have
more than 10 space points in the VTPCs. Short tracks were eliminated by
requiring that the track segment in the MTPC was longer than 1.5~m
in order to obtain good \dedx~measurements and minimize the effect of track splitting.
Additionally, tracks were required to have a good quality trajectory fit.
To guarantee precise time measurements and reject tracks depositing
too little energy in the tiles because of the edge effect, a cut on the energy
deposited in a scintillator was imposed discarding the lowest 10\% of the pulse
height distribution in a tile.
Moreover, if more than one MTPC track candidate was matched to the same
scintillator tile, resulting in an ambiguous time-of-flight measurement,
these tracks were removed from the analysis.   

\subsection{Identification of light nuclei}
The identification of light nuclei ($d,t,^3$He) was based on 
momentum, \dedx, and time-of-flight measurements.
Deuteron and triton candidates were required to have a TOF hit matched
to the MTPC track.
\begin{figure}
\begin{center}
\begin{minipage}{17pc}
\includegraphics[width=17pc]{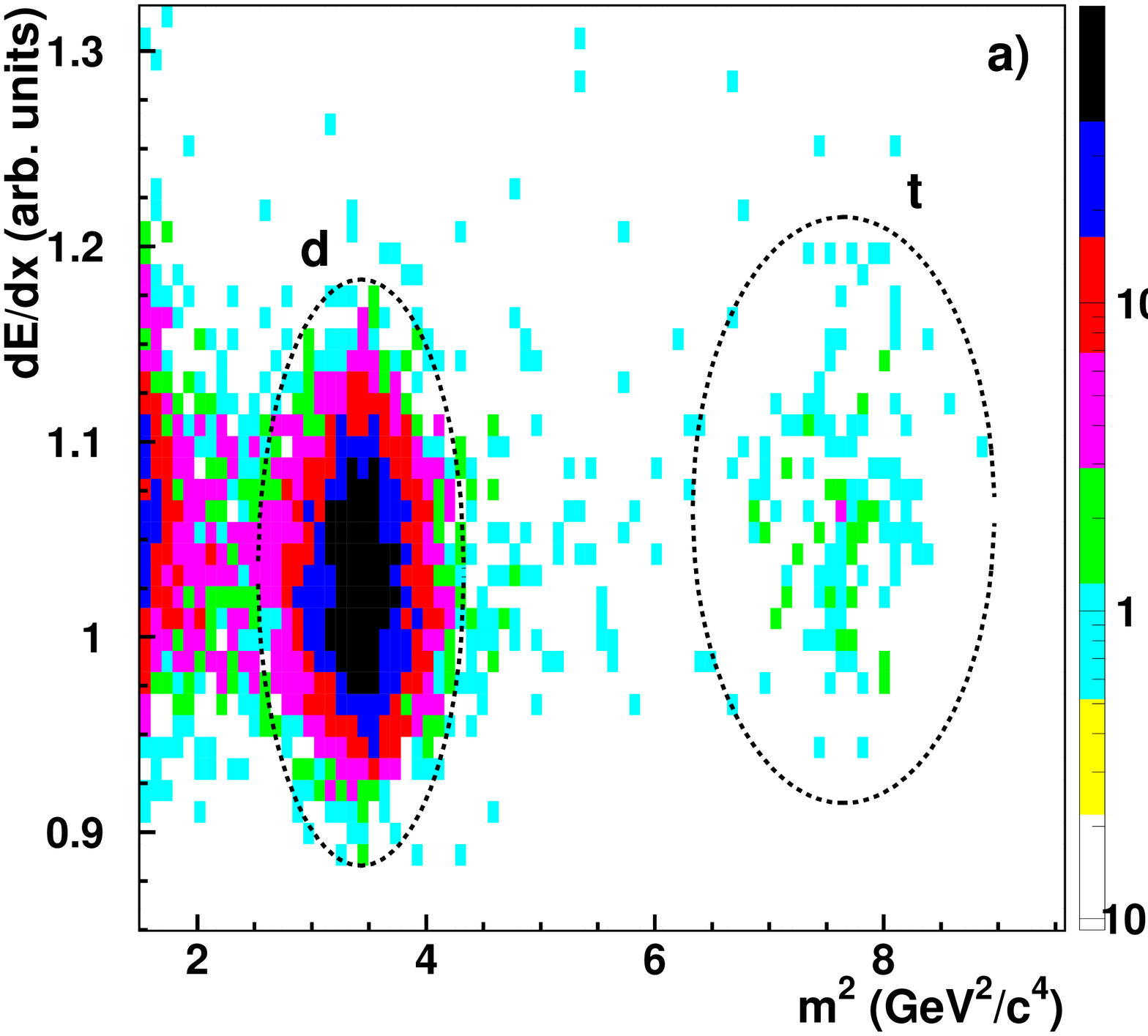}
\end{minipage}
\begin{minipage}{17pc}
\includegraphics[width=17pc]{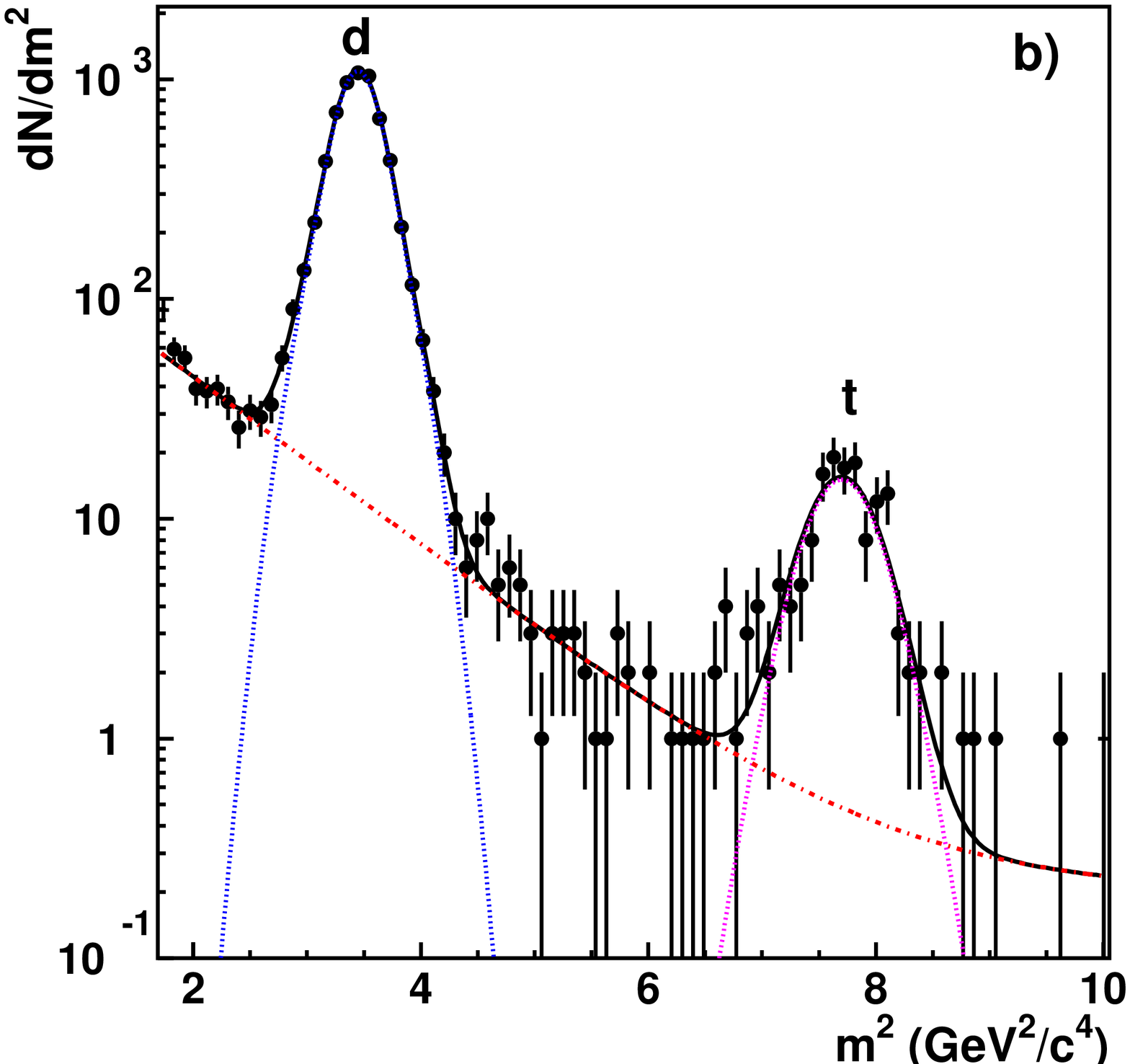}
\end{minipage}\hspace{0pc}%
\caption{\label{fig1}(Color online) a) Energy loss \dedx~versus mass-squared from Pb+Pb
at 20\agev~for the momentum interval 6$<$p$<$8~\gevc~. Deuteron and triton
candidates are selected within the 3$\sigma$ PID ellipses indicated by the dashed lines.  
b) Mass-squared distribution after a \dedx~upper limit cut (see text). The solid line indicates
the best fit with two Gaussians (the dotted lines) and the background (the dash-dotted line).}
\end{center}
\end{figure}
Identification of deuterons and tritons was performed in momentum bins of 2~\gevc~ width
by selecting particles with measured values of \dedx~and $m^2$
within three standard deviations of the expected values (see Fig.~\ref{fig1}(a)). 
The background contamination in both the deuteron
and triton samples was estimated by analyzing the projection of the
\dedx~versus $m^2$ histogram  onto the $m^2$-axis with an upper limit \dedx~cut
applied: $\dedx<$~($\langle \dedx \rangle_{t} +3\sigma_{t})$, where
$\langle \dedx \rangle_{t}$ is the predicted value for tritons and $\sigma_t$ is the
\dedx~resolution.  The obtained distribution (see example in Fig.~\ref{fig1}(b))
consisting of two signal peaks for $d$ and $t$ plus some background was then fitted
to a sum of two Gaussians superimposed on an exponential plus first-order polynomial function.
The raw yield was calculated by counting the entries in mass windows
of 2.5$<$$m^2$$<$4.5 and 6.9$<$$m^2$$<$7.9~GeV$^2$/$c^4$ for $d$ and $t$, respectively.
The percentage of counts outside of the mass window was
estimated from the Gaussian signal shape;  the background contribution, not exceeding
10\% within the studied momentum range, was subtracted from the data.

\begin{figure}
\includegraphics[width=1.0\linewidth]{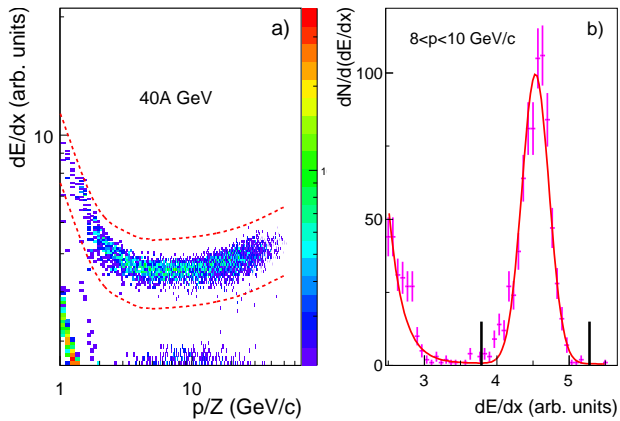}
\caption{(Color online) a) Energy loss \dedx~measured in the MTPC as a
function of rigidity for charged tracks from central Pb+Pb collisions at 40\agev.
The dashed curves indicate the PID cut boundaries used for selection
of $^3$He. b) Distribution of \dedx~in the momentum interval $8<p<10$~\gevc.
A fit of a Gaussian signal plus background is shown by the solid
curve. Black vertical lines indicate the selection window for $^3$He.}
\label{fig3_2}
\end{figure}

The identification of $^3$He candidates can rely completely on the specific energy loss
measurement in the MTPC gas on account of their double charge.
Since matching to TOF is not required the kinematic acceptance is much larger than
for deuterons and tritons.
Due to overlap of the \dedx~bands for $^3$He and $^4$He at
momenta above 10~\gevc, the latter species contaminates the $^3$He selection band.
Based on the scaling behavior of light nuclei yields with increasing mass
number $A$ (see Section IVB, Eq.~\ref{penalty}), the $^4$He contamination in the selected candidates 
was, however, estimated to be below 3\% at all collision energies and was neglected. 
Helium nuclei were selected by a $3\sigma$ cut around the predicted \dedx~position 
as indicated in the example shown in Fig.~\ref{fig3_2}(a). In order to estimate
the background of misidentified particles in the $^3$He samples
the distributions were projected onto the \dedx~axis in bins of momentum.
The projected distributions were then fitted by a Gaussian for the signal plus a sum
of a first-order polynomial and an exponential function for the background
(see an example in Fig.~\ref{fig3_2}(b)).   
%As can be seen in the sample PID plot shown in Fig.~\ref{fig3_2} (a),
%clear discrimination between He nuclei and other particles is possible 
%at all measured momenta. The distributions were projected onto the \dedx~axis.
%Helium nuclei were selected by a $3\sigma$ cut around the predicted \dedx~position 
%as indicated in the example shown in Fig.~\ref{fig3_2} (b).
%The projected distributions were fitted by a Gaussian for the signal plus a sum
%of a first-order polynomial and an exponential function for the background.
The background contribution, varying with beam energy between 2\% at
20\agev and 20\% at 158\agev, was subtracted from the data.

\subsection{Corrections}
After selecting the $d$, $t$, and $^3$He samples, all light nuclei candidates
are binned in rapidity $y$, transverse momentum $p_t$, and transverse
mass $m_t$$-m$ defined as:
$$
y=\frac{1}{2}\ln{\frac{E+p_z}{E-p_z}},~~p_t=\sqrt{p_x^2+p_y^2},~~m_t=\sqrt{p_t^2+m^2},
$$
where $E$ and $p_z$ are the energy and longitudinal momentum component in the
center-of-mass system, $p_t$, $p_x$, $p_y$ the transverse momentum components and
$m$ the rest mass of the candidate.
The raw yields of clusters need to be corrected for geometrical acceptance,
detector efficiency, and for the losses due to the applied cuts and PID selection
criteria.

\begin{figure}
\includegraphics[width=1.0\linewidth]{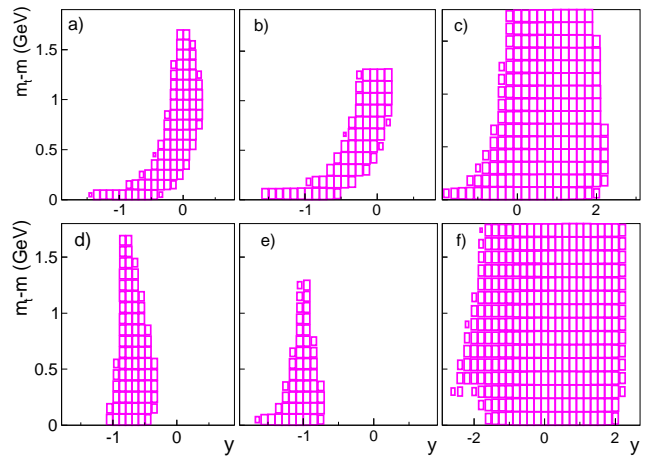}
\caption{(Color online) The NA49 phase space coverage in terms of $y$ and
$m_t$ at 20\agev~for deuterons (a), tritons (b), $^3$He (c), and
at 158\agev~for deuterons (d), tritons (e), $^3$He (f).}
\label{acc_all}
\end{figure}

The correction for the limited geometrical acceptance was obtained from Monte
Carlo (MC) simulations. In order to have similar statistics of simulated tracks
in different phase space bins, MC tracks were taken from a flat distribution
over momentum $p$, polar angle $\theta$, and azi\-mu\-thal angle $\phi$.
The particles were propagated through the detector setup with the GEANT3
package to determine the measurable track length and the potential number
of space points. Tracks were then checked to satisfy all the geometrical fiducial cuts and number of
space point cuts imposed on real data. Ionization energy loss and multiple scattering 
were taken into account by the GEANT3 tracking. Dead or inefficient TOF tiles not used
in data analysis were removed from simulations.
An acceptance map was generated for each cluster species and at each beam
energy (i.e. magnetic field setting). Figure~\ref{acc_all} shows the NA49
phase space coverage for $d, t$ and $^3$He in terms of transverse mass and rapidity
at the lowest and the highest beam energy.

The tracking efficiency was studied by embedding simulated tracks into
real data; it was found to be close to 100\%. All the corrections due to quality
cuts were evaluated from the data. The inefficiency due to multiple tracks hitting
the same TOF tile varies from 6\% (20\agev) to 11\% (158\agev); the
corresponding corrections were obtained by counting rejected tracks.
The correction for the TOF hit pulse height cut is typically 10\% and is 
weakly dependent on the beam momentum and particle species. Corrections for
absorption along the trajectory through the detector material were, however, not taken
into account since nucleus-nucleus interactions are not well described in
GEANT3.
%added 25.05.16 V.I. Kolesnikov
Nevertheless, a rough estimate of the losses due to inelastic interactions of $d$,
$t$, and $^3$He in the material can be made based on the si\-mu\-la\-ted absorption of protons.
The latter was obtained by switching on and off nuclear interactions in GEANT3
and then comparing the fraction of protons that survived when passing
through the detector reaching the TOF wall. Because the momenta of protons registered
in the TOF wall is above 1 GeV/c, the absorption loss for protons having a TOF hit
was found to be weakly dependent on rapidity and $p_t$ and did not exceed 3.5\%.
Making an assumption on how the inelastic cross-section scales with atomic mass
number $A$~\cite{cross_sections}, the upper limit of the absorption loss was
estimated to be of 5\% and 7\% for $A$=2 and $A$=3 nuclei, respectively. 
The estimated uncertainty associated with the absorption losses enters as a contribution
to the overall systematic error.
  
Since the secondary nuclei knocked out by hadronic interactions in the material
have momenta considerably lower than the low-$p$ cut-off of the TOF acceptance
of 1~\gevc, their contribution in the analyzed data samples is negligible. 
%end of adding

\subsection{Systematic uncertainties}
\label{errors}
The main sources of the overall systematic uncertainty originate 
from extraction of the raw yields (including background subtraction) and from
efficiency correction factors. Each particular contribution to the systematic
uncertainty associated with the extraction procedure as well as with corrections
due to the applied PID and qua\-li\-ty cuts (described in the previous section) was
estimated by varying one-by-one the respective selection criteria and repeating
the analysis procedure.
Thus, the uncertainty of background subtraction was estimated by varying the value
of the \dedx~PID cut and changing the fit range and functional shapes for the
background: the resulting raw yields were found to be consistent within 3\%
for $d$ and $^3$He and 5\% for $t$, respectively. The error related to the pulse
height cut is estimated at 1-2\% (depending on the beam momentum). An
estimate for the systematic uncertainty due to the TOF multi-hit cut was obtained from
the spread of the difference  of individual tile-wise multi-hit corrections
with respect to the one ave\-ra\-ged over all the scintillators. The spread
was found to vary within 2\% over the considered $p_t$ range and slightly depended
on the collision energy.
The various contributions, including uncertainties associated with the absorption losses,
were added quadratically resulting in a total
systematic uncertainty of 6\% for deuterons and 9\% for tritons and $^3$He. 

In order to investigate the reliability of these estimates, the datasets at 40$A$ and
158\agev~were divided into two parts that were taken with opposite directions
of the magnetic field in the VTPCs (named as $STD+$ and $STD-$ setting).
At opposite polarities of the field the produced charged particles are
registered in the opposite halves (relative to the beam line)
of the detector. The $^3$He analysis was then performed in each data subset
se\-pa\-ra\-te\-ly and the difference in the fiducial yields averaged
over all rapidity bins was of order 8\% and 10\%
at 40$A$ and 158\agev, respectively. As the final yields are the 
average for the $STD+$ and $STD-$ data sets the results agree
within the uncertainties estimated above.

\section{Results and discussion}

\subsection{Transverse momentum spectra and yields}

\begin{figure*}[t]
\begin{minipage}[b]{0.95\linewidth}
\includegraphics[width=0.95\linewidth]{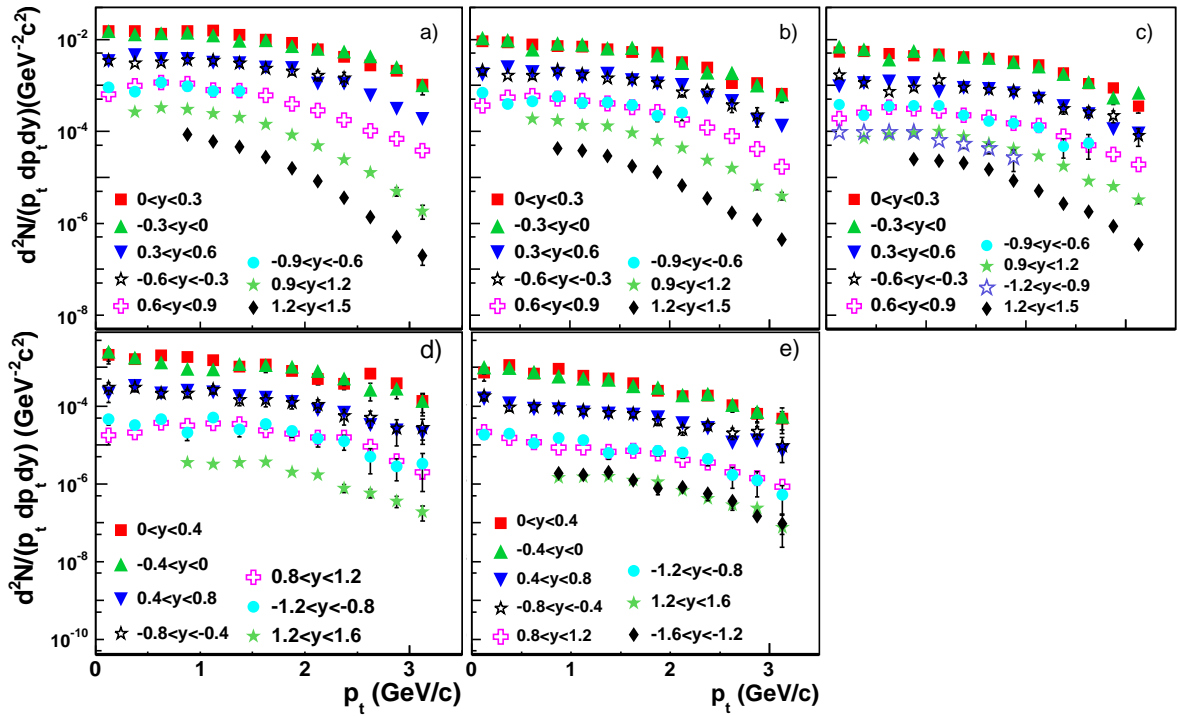}
\end{minipage}
\caption{(Color online) Invariant $p_t$ spectra of $^3$He at 20$A$ (a),
30$A$ (b), 40$A$ (c), 80$A$ (d), and 158\agev~(e). Only statistical errors are
shown. The distributions near mid-rapidity are drawn to scale, other
spectra are scaled down by successive powers of 5 for clarity. The same
scaling factor is used for two rapidity slices that are symmetric about
mid-rapidity ($y=0$).}
\label{all_pt_hel}
\end{figure*}

%We first discuss the transverse momentum distributions of light nuclei.
The invariant $p_t$ spectra of identified $^3$He nuclei at five collision energies
in different rapidity intervals are shown in Fig.~\ref{all_pt_hel}.
The interval sizes range from 0.3 at 20$A$-40\agev~to 0.4 at 80$A$ and 158\agev.
The expe\-ri\-men\-tal distributions were scaled down successively
for clarity of presentation. The same scaling factors were used for the rapidity
slices located symmetrically with respect to the center of mass.
First a test was performed whether the yields in the rapidity bins symmetric relative
to $y = 0$ are consistent. The test procedure includes the following
steps. The distribution at a particular forward rapidity bin (source) was first fitted with
an appropriate function: a sum of two exponential functions was employed.
Then the fit parameters defining the shape of the spectrum were fixed and the
"mirrored" spectrum (target) at corresponding backward rapidity was fitted with only the normalization
parameter allowed to vary. Finally, $\chi^2$ per point for the target spectrum with
respect to the shape of the source distribution was determined within the common
$p_t$ acceptance range. The procedure was then repeated switching the source
and target spectrum. The result was that for all colliding
energies and rapidity intervals the $\chi^2/$NDF values ranged from 0.4 to 1.9, with
rapidity averaged values $\langle\chi^2$/NDF$\rangle$ of 1.3, 1.2, 1.6, 1.1,
and 1.2 at 20$A$, 30$A$, 40$A$, 80$A$ and 158\agev, respectively.
Such a low value of the $\chi^2$ per degree of freedom indicates that the results from forward
rapidities are consistent within their uncertainties with those at backward rapidities.
%Thus, analyzing the changes of the spectra shapes with rapidity we combined
%the data for $^3$He from two symmetric around mid-rapidity slices.

\begin{figure*}
\begin{minipage}[b]{0.95\linewidth}
\includegraphics[width=0.95\linewidth]{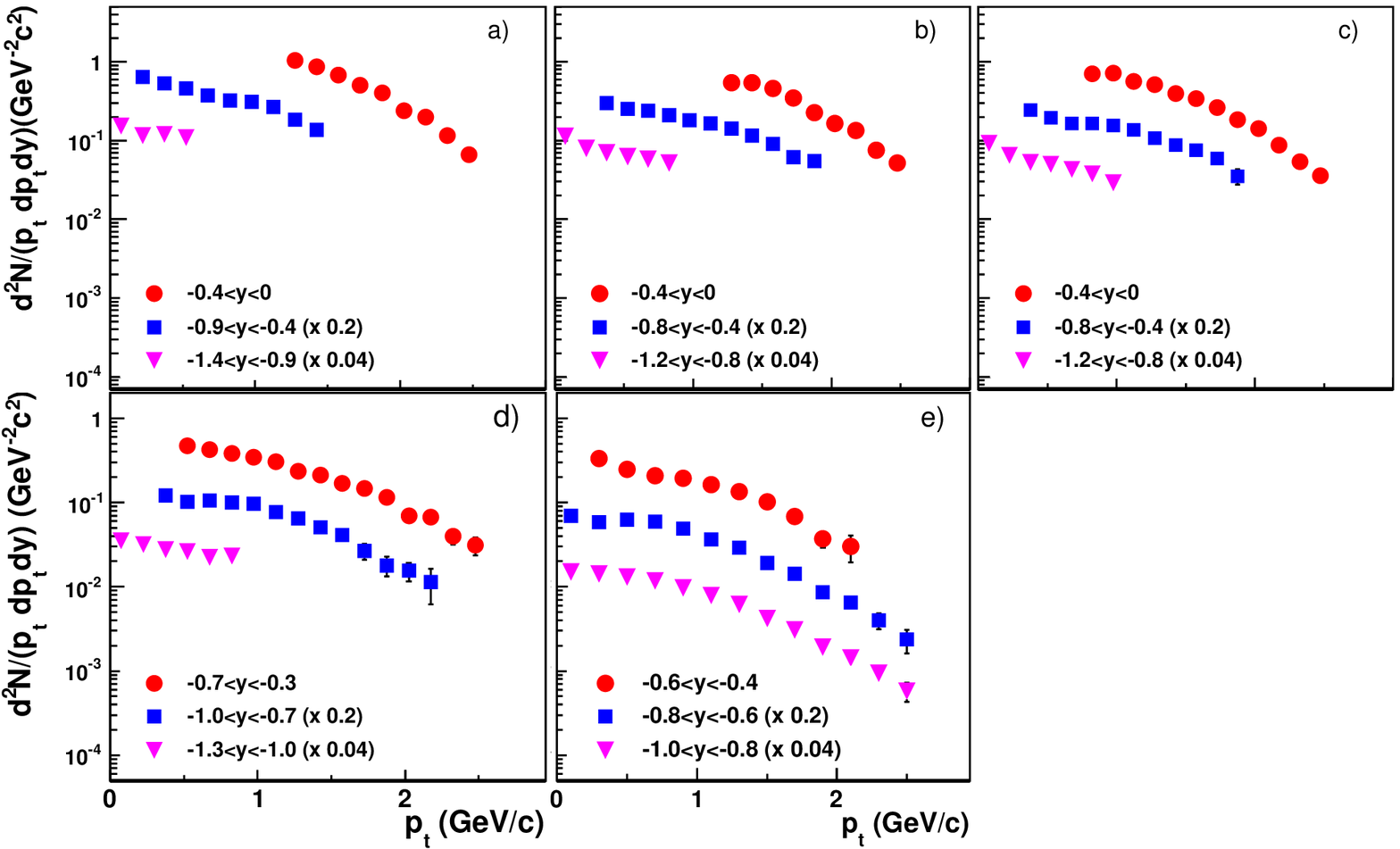}
\end{minipage}
\caption{(Color online) Invariant $p_t$ spectra of $d$ at 20$A$ (a),
30$A$ (b), 40$A$ (c), 80$A$ (d), and 158\agev~(e). Only statistical errors are
shown.}
\label{all_pt_deu}
\end{figure*}

The NA49 acceptance for deuterons is sufficient to exa\-mi\-ne $p_t$ spectra
of $d$ in three rapidity intervals. Fi\-gu\-re~\ref{all_pt_deu} presents the
corresponding invariant $p_t$ spectra of deuterons at five collision energies.
The intervals are specified in the legends of the figure. 

\begin{figure}
\includegraphics[width=1.0\linewidth]{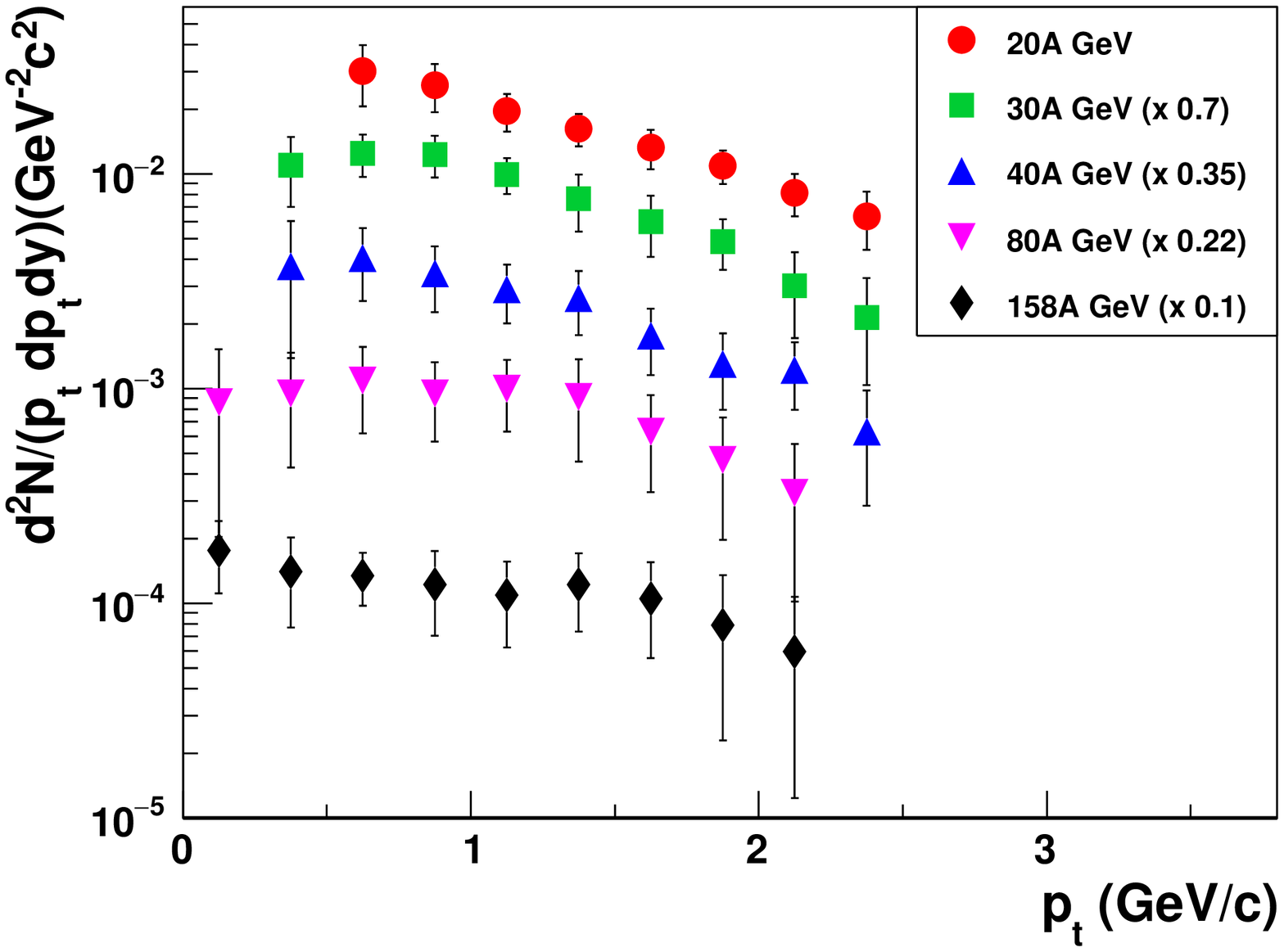}
\caption{(Color online) Invariant $p_t$ spectra of $t$ at 20$A$-158\agev.
Only statistical errors are shown.}
\label{all_pt_tr}
\end{figure}

The statistics for tritons is, however, too low to obtain
meaningful $p_t$ spectra in several rapidity bins.
Thus, for $t$ the results for each $p_t$ bin were integrated over the TOF rapidity acceptance.
Figure~\ref{all_pt_tr} shows the invariant yield of tritons versus
$p_t$ at all bombarding energies. 
 
\begin{figure}
\includegraphics[width=0.9\linewidth]{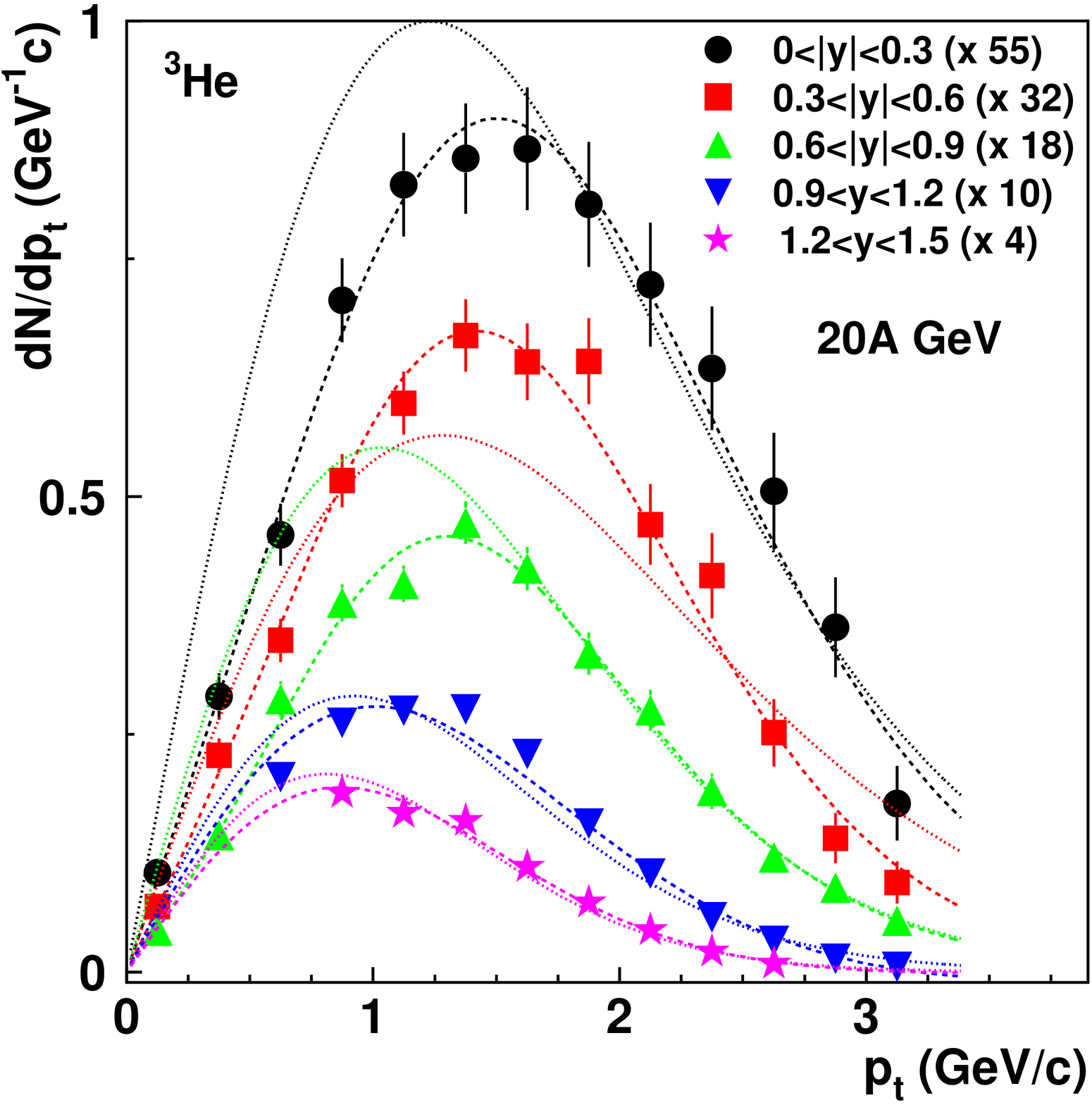}
\caption{(Color online) $p_t$ spectra of $^3$He in rapidity bins
of $|\Delta y|=0.3$ from central Pb+Pb collisions at 20\agev.
Fits with a sum of two exponentials are plotted by the dashed curves, fits with a single exponential
by dotted curves (see text for more detail).}
\label{all_pt_20_lin}
\end{figure}

In order to obtain \dndy~the measured $p_t$ distributions
need to be extrapolated into unmeasured $p_t$ regions exploiting
information on the spectral shape.
For this, $^3$He spectra were first tested with an exponential function:
\begin{equation}
\label{eq:b1}
\frac{d^2N}{dp_tdy}= \frac{\dndy}{T(m+T)} \, p_t \, \exp{\left(-\frac{m_t-m}{T}\right)}~~,
\end{equation}
where \dndy~and $T$ are two fit parameters, $m_t = \sqrt{p_t^2 + m^2}$
is the transverse mass and $m$ is the $^3$He rest mass. 
Such a functional form reproduces most meson spectra from heavy-ion
collisions quite well~\cite{e802_ka,na49_pika}. However, the description
of the shapes of the mid-rapidity spectra of light nuclei by Eq.~\ref{eq:b1} is not satisfactory.
As can be seen in Fig.~\ref{all_pt_20_lin}, single-exponential fits (plotted
with dotted lines) overestimate  mid-rapidity $p_t$ spectra of $^3$He at low and
high transverse momenta. The degree of deviation is indicated by a typical
$\chi^2/NDF$ of about 7.
A sum of two exponentials describes the mid-rapidity spectra
much better (see dashed lines in Fig.~\ref{all_pt_20_lin}).
Thus such a parameterization was used for extrapolation of the $p_t$ spectra of light nuclei
with respect to mid-rapidity. The observed difference between the two- and single-exponential
fits, however,  diminishes towards forward rapidities (see the results for blue down-pointing triangles and pink stars in Fig.~\ref{all_pt_20_lin}).
Since the single-exponential function of Eq.~\ref{eq:b1}
produces much more stable fit results for spectra with limited $p_t$ coverage at low
transverse momenta it was used for the extrapolation
of spectra at very forward rapidities down to $p_t=0$.
The extrapolation amounts to 3-7\% of the \dndy~value
for the $^3$He  spectra near mid-rapidity and increases to almost
40\% for the results at the most forward rapidity. 

For the case of deuterons the spectra from two adjacent
rapidity bins were combined to obtain the pa\-ra\-me\-ters defining the spectral shape
from a fit to a sum of two exponentials. These parameters were then fixed for the extrapolation
of each spectrum from the combination to the unmeasured region.
The extrapolation for the not-covered $p_t$ region is less than 10\%
at 158\agev, the amount of extrapolation at lower energies varies from 5 to 25\%
at mid-rapidity and increases up to 70\% for the most backward rapidity
bin at 20\agev.

\begin{figure*}
\begin{minipage}[b]{0.95\linewidth}
\includegraphics[width=0.95\linewidth]{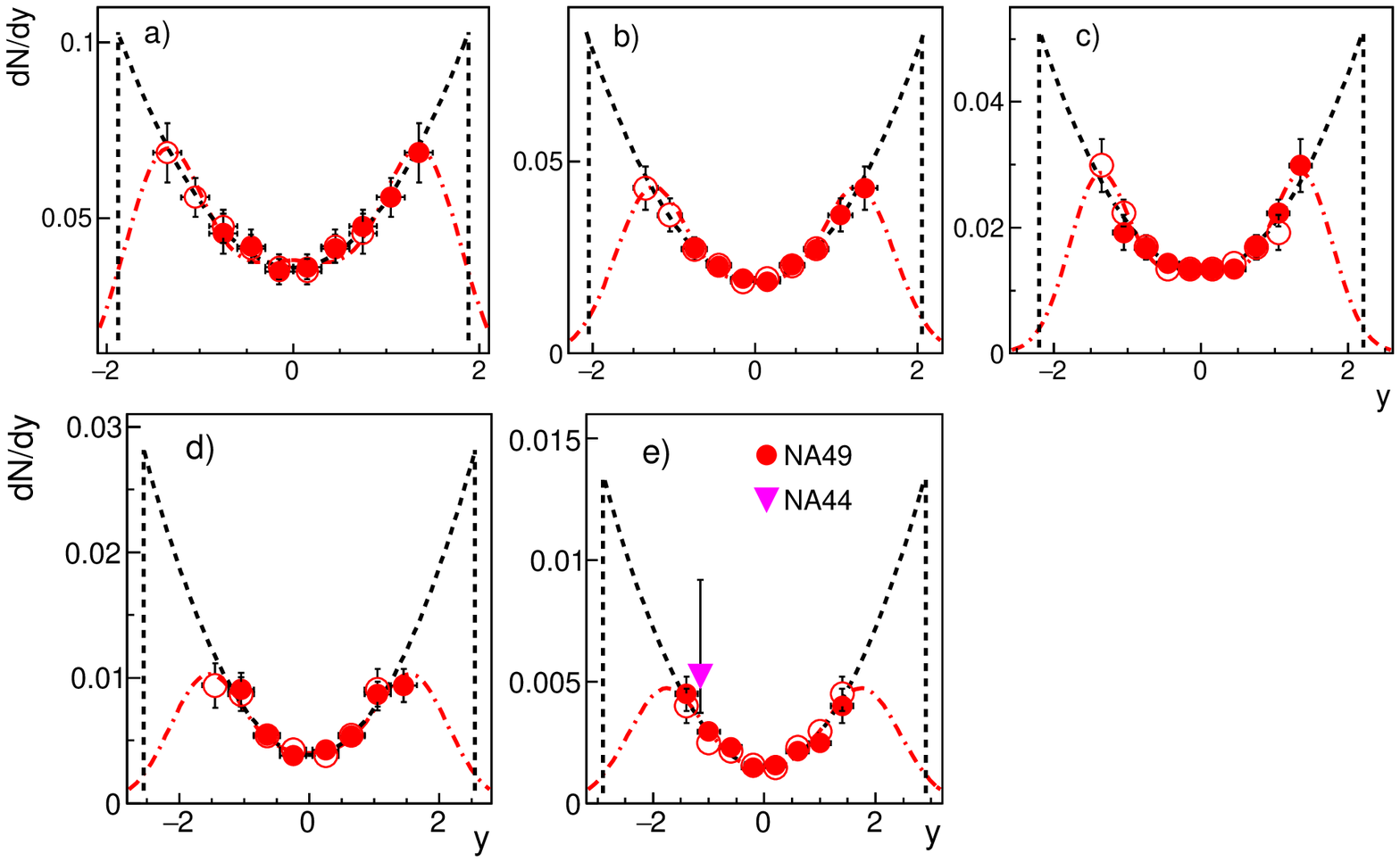}
\end{minipage}
\caption{(Color online) Rapidity distributions for $^3$He at 20$A$ (a),
30$A$ (b), 40$A$ (c), 80$A$ (d), and 158\agev~(e). The solid symbols show the
measurements and the open symbols represent the data points
reflected about mid-rapidity. The error bars correspond to the quadratic
sum of statistical and systematic errors. Dashed and dot-dashed lines indicate the functional
forms used to extrapolate to $4\pi$ yields (see text for more detail).
The NA44 experimental data on $t$ are taken from Ref.~\cite{deu_na44}}
\label{dndy_hel}
\end{figure*}

\begin{figure*}
\begin{minipage}[b]{0.95\linewidth}
\includegraphics[width=0.95\linewidth]{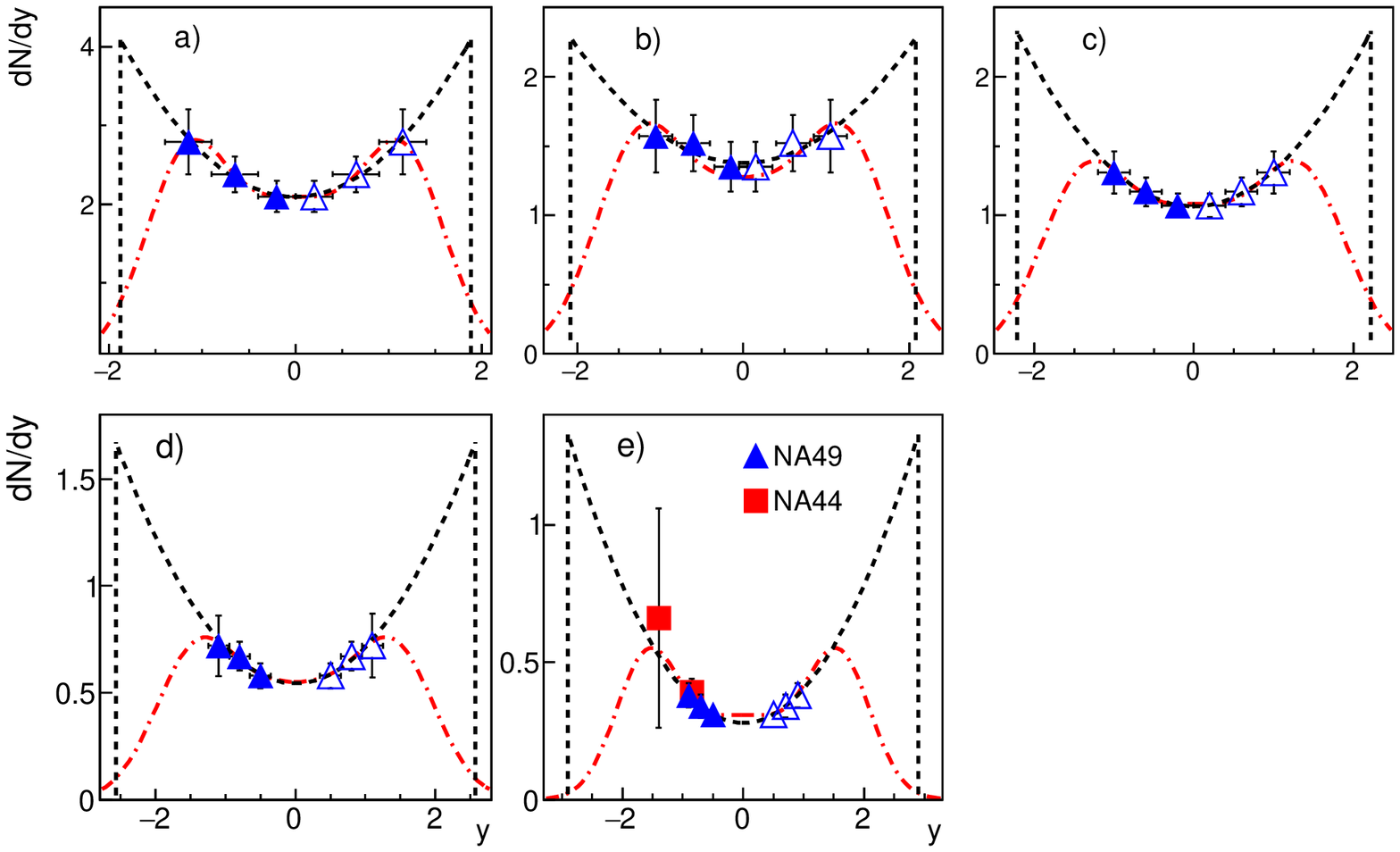}
\end{minipage}
\caption{(Color online) Rapidity distributions for $d$ at 20$A$ (a),
30$A$ (b), 40$A$ (c), 80$A$ (d), and 158\agev~(e). The solid symbols show the
measurements and the open symbols represent the data points
reflected about mid-rapidity. The error bars correspond to the quadratic
sum of statistical and systematic errors. Dashed and dot-dashed lines indicate the functional
forms used to extrapolate to $4\pi$ yields (see text for more detail).
The NA44 experimental data on $d$ are taken from Ref~\cite{deu_na44}.}
\label{dndy_deu}
\end{figure*}

The systematic uncertainty of \dndy~arises from the uncertainty of spectra
normalization and of the ex\-tra\-po\-la\-tion procedure.
Regarding the first contribution, the overall systematic uncertainty for the yields
of clusters was estimated to 6-9\% (see Section~\ref{errors}).

The systematic uncertainty of \dndy~for the data with a limited
$p_t$ coverage  is largely determined by the extrapolation procedure.
The uncertainty associated with the extrapolation was estimated by using different
functions: single exponential, sum of two exponentials and Boltzmann form.
It was found that the difference in the results for the extrapolation using
different fit functions for $t$ is about 7\%. For $^3$He the method gives
a typical uncertainty of 1-3\% at mid-rapidity and approximately 10\% for the most
forward rapidity bin. For deuterons this systematic uncertainty varies from 5\% to 15\%.

\begin{table}
\caption{\label{table:yields}The yield \dndy~of $^3$He in rapidity slices $(y_1,y_2)$}
\begin{ruledtabular}
\begin{tabular}{llll}
 $(y_1,y_2)$& $\dndy\cdot10^{3}$ &$(y_1,y_2)$ & $\dndy\cdot10^{3}$ \\\hline
\multicolumn{4}{c}{20\agev}\\
$(-0.9,-0.6)$ & 45.5 \tpm 5.7 & $(0.3,0.6)$ & 41.2 \tpm 4.1\\
$(-0.6,-0.3)$ & 41.9 \tpm 4.7 & $(0.6,0.9)$ & 47.6 \tpm 4.8 \\
$(-0.3,0.0)$ & 34.7 \tpm 3.6 & $(0.9,1.2)$ & 55.9 \tpm 5.6 \\
$(0.0,0.30)$ & 36.0 \tpm 3.6 & $(1.2,1.5)$ & 68.6 \tpm 8.5 \\
\multicolumn{4}{c}{30\agev}\\
$(-0.9,-0.6)$ & 27.2 \tpm 3.0 & $(0.3,0.6)$ & 23.1 \tpm 2.3 \\
$(-0.6,-0.3)$ & 22.8 \tpm 2.4 & $(0.6,0.9)$ & 27.1 \tpm 2.7 \\
$(-0.3,0.0)$ & 19.4 \tpm 2.0 & $(0.9,1.2)$ & 36.1 \tpm 4.0 \\
$(0.0,0.3)$ & 18.7 \tpm 2.0 & $(1.2,1.5)$ & 43.1 \tpm 5.7 \\
\multicolumn{4}{c}{40\agev}\\
$(-1.2,-0.9)$ & 19.2 \tpm 2.8 & $(0.3,0.6)$ & 13.3 \tpm 1.4 \\
$(-0.9,-0.6)$ & 16.8 \tpm 1.9 & $(0.6,0.9)$ & 17.0 \tpm 1.7  \\
$(-0.6,-0.3)$ & 14.3 \tpm 1.6 & $(0.9,1.2)$ & 22.3 \tpm 2.2\\ 
$(-0.3,0.0)$ & 13.4 \tpm 1.4 & $(1.2,1.5)$ & 29.9 \tpm 4.2 \\
$(0.0,0.3)$ & 13.4 \tpm 1.4 \\
\multicolumn{4}{c}{80\agev}\\
$(-1.25,-0.85)$ & 9.1 \tpm 1.3 & $(0.45,0.85)$ & 5.4 \tpm 0.6 \\
$(-0.85,-0.45)$ & 5.4 \tpm 0.7 & $(0.85,1.25)$ & 8.7 \tpm 1.0 \\
$(-0.45,-0.05)$ & 3.8 \tpm 0.5 & $(1.25,1.65)$ & 9.4 \tpm 1.3 \\
$(0.05,0.45)$ & 4.3 \tpm 0.5\\
\multicolumn{4}{c}{158\agev}\\
$(-1.6,-1.2)$ & 4.5 \tpm 0.7 & $(0.0,0.4)$ & 1.6 \tpm 0.2 \\
$(-1.2,-0.8)$ & 2.9 \tpm 0.4 & $(0.4,0.8)$ & 2.1 \tpm 0.3\\
$(-0.8,-0.4)$ & 2.3 \tpm 0.4 & $(0.8,1.2)$ & 2.5 \tpm 0.3 \\
$(-0.4,0.0)$ & 1.5 \tpm 0.2 & $(1.2,1.6)$ & 4.0 \tpm 0.7\\
\end{tabular}
\end{ruledtabular}
\end{table}

\begin{table}
\caption{\label{table:yields1}The yield \dndy~of $d$ in rapidity slices ($y_1,y_2$)}
\begin{ruledtabular}
\begin{tabular}{llll}
$(y_1,y_2)$ & \dndy & $(y_1,y_2)$ & \dndy \\\hline
\multicolumn{4}{c}{20\agev}\\
$(-1.4,-0.9)$ & 2.79 \tpm 0.44 & $(-0.4,0.0)$ & 2.10 \tpm 0.22 \\
$(-0.9,-0.4)$ & 2.38 \tpm 0.22 & &\\
\multicolumn{4}{c}{30\agev}\\
$(-1.2,-0.8)$ & 1.57 \tpm 0.26 & $(-0.4,0.0)$ & 1.35 \tpm 0.18 \\
$(-0.8,-0.4)$ & 1.52 \tpm 0.20 & &\\
\multicolumn{4}{c}{40\agev}\\
$(-1.2,-0.8)$ & 1.31 \tpm 0.16 & $(-0.4,0.0)$ & 1.07 \tpm 0.11\\ 
$(-0.8,-0.4)$ & 1.17 \tpm 0.12 & &\\
\multicolumn{4}{c}{80\agev}\\
$(1.3,-1.0)$ & 0.72 \tpm 0.13 & $(-0.6,-0.2)$ & 0.58 \tpm 0.06\\
$(-1.0,-0.6)$ & 0.67 \tpm 0.07 & &\\
\multicolumn{4}{c}{158\agev}\\
$(-1.0,-0.8)$ & 0.38 \tpm 0.04 & $(-0.6,-0.4)$ & 0.31 \tpm 0.04 \\
$(-0.8,-0.6)$ & 0.34 \tpm 0.04 & &\\
\end{tabular}
\end{ruledtabular}
\end{table}

The results on \dndy~are tabulated in Table~\ref{table:yields}
for $^3$He and in Table~\ref{table:yields1} for deuterons. The quoted
total uncertainties are the quadratic sums of the statistical and systematic uncertainties.
Figures~\ref{dndy_hel} and~\ref{dndy_deu} present the yields of $^3$He
and $d$ at all beam momenta as a function of rapidity. Measurements
are plotted by solid symbols and open points show reflections of measurements
around mid-rapidity.

The NA44 experiment studied the production of deuterons and tritons
in central Pb+Pb interactions at the top SPS energy.  Their experimental
data~\cite{deu_na44} on $d$ in the 10\% and $t$ in the 20\% most central
Pb+Pb collisions are plotted along with the present measurements in Fig.~\ref{dndy_deu}(e)
and Fig.~\ref{dndy_hel}(e), respectively. Although centrality selections
differ slightly and the yield of tritons is somewhat higher than that
of $^3$He at SPS energies (see Section~\ref{trhel_rat}), the agreement
between the two experiments can be considered reasonable.

In order to extrapolate the integral of \dndy~to
full phase space two different parameterizations of the rapidity spectra
were employed which provide lower and upper limits of the
integral. For the lower limit the same parameterization was used
as in Ref.~\cite{dbar_na49}. There it was found that a sum
of three Gaussians (one centered at mid-rapidity and two others
displaced symmetrically relative to $y=0$) describes the rapidity spectrum
of deuterons from mid-central Pb+Pb collisions at 158\agev~quite well.
This picture is based on the assumption that the observed particle emission
pattern requires (at least) three sources: two located close to the
(quasi)projectile/target rapidity and one at mid-rapidity. 
Fits using this function ('Fit A') are indicated in Figs.~\ref{dndy_hel}~and~\ref{dndy_deu}
by dot-dashed red lines. 
Extrapolations obtaining the upper limit ('Fit B') are based on the assumption that
the longitudinal freezeout distribution spans the entire rapidity range with
a broad minimum at mid-rapidity.
This behavior was parameterized by a parabolic function with a sharp drop at $\pm y_{beam}$
(see black dashed histograms in Figs.~\ref{dndy_hel}~and~\ref{dndy_deu}).
The total yield was then obtained by summing the measured values with the integral
of the corresponding extrapolation function over the unmeasured region.
The extrapolation accounts from 30\% to 63\% and from 20\% to 85\% of the
4$\pi$ yield for $^3$He and $d$, respectively, depending on the collision
energy and type of extrapolation ('Fit A' or 'Fit B').
\begin{figure}
\includegraphics[width=1.0\linewidth]{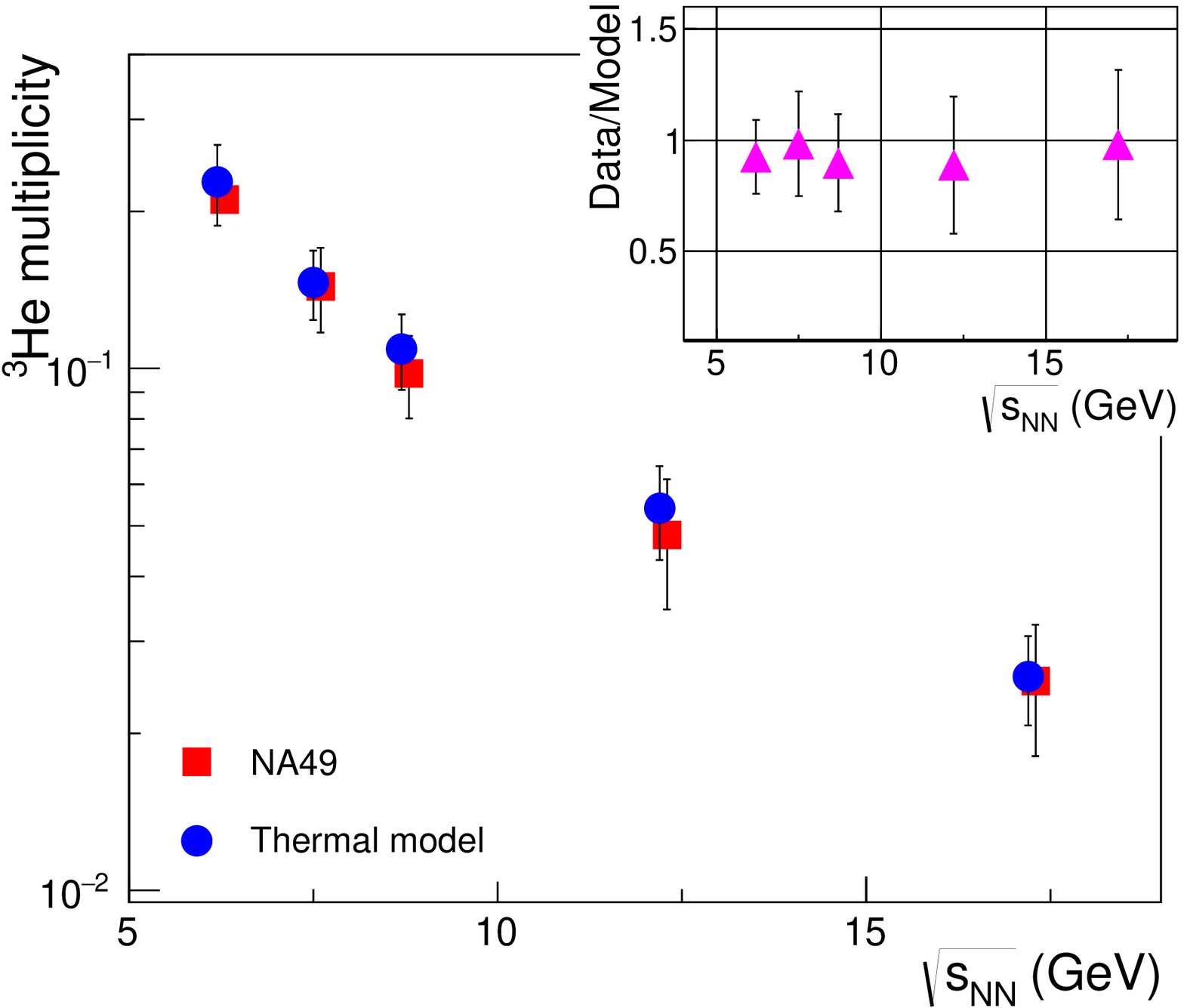}
\caption{(Color online) 4$\pi$ yield of $^3$He
in central Pb+Pb collisions at 20$A$-158\agev. The NA49 data (red squares)
are the average of the results for the 'Fit A' and 'Fit B' extrapolations
(see text for detail).
Thermal model calculations (see text) are shown by blue circles;
the inset shows the ratio of the experimental data to the thermal model predictions.
Symbols for experimental data and model predictions have been displaced
for clarity in presentation.}
\label{helium_4pi}
\end{figure}

\begin{figure}
\includegraphics[width=1.0\linewidth]{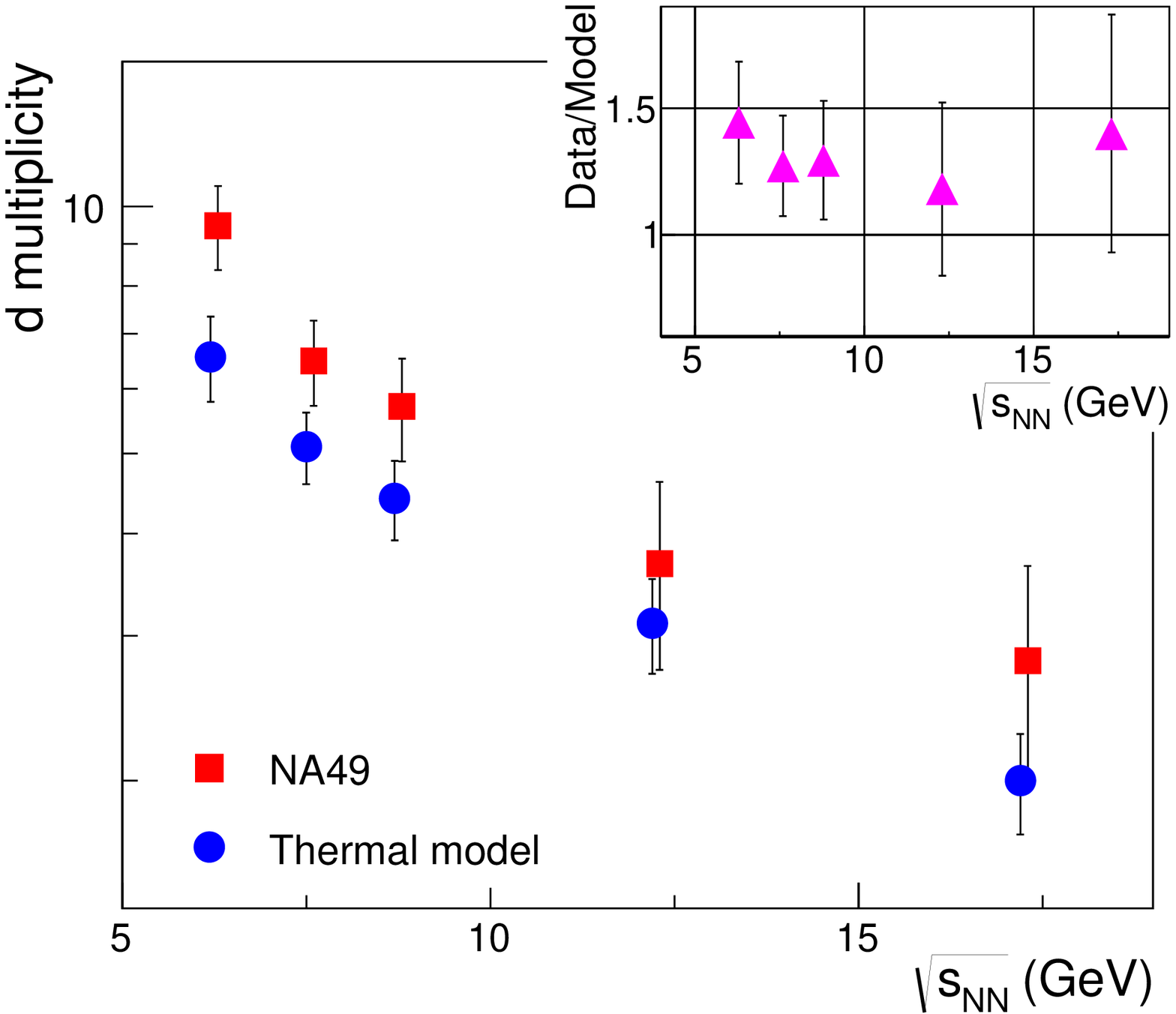}
\caption{(Color online) 4$\pi$ yield of deuterons
in central Pb+Pb collisions at 20$A$-158\agev. The NA49 data (red squares)
are the average of the results for the 'Fit A' and 'Fit B' extrapolations (see
text for detail). Thermal model calculations (see text) are shown by blue circles;
the inset shows the ratio of the experimental data to the thermal model
predictions. Symbols for experimental data and model predictions have been
displaced for clarity in presentation.}
\label{deu_4pi}
\end{figure}

The resulting estimates for the total yields (multiplicities) of $^3$He and deuterons
are tabulated in Table~\ref{table:total} for the two extrapolation functions discussed above.
The average between these two estimates is plotted
versus $\sqrt{s_{NN}}$ in Figs.~\ref{helium_4pi}~and~\ref{deu_4pi}
for $^3$He and $d$, respectively. The plotted overall uncertainty for the mean is a
combination of the squares of the data point uncertainties and the extrapolation
uncertainty caused by the lack of knowledge about the true shapes of the \dndy~
distributions near the beam(target) rapidity. The latter uncertainty was estimated
as half of the difference between the 'Fit A' and 'Fit B' extrapolations
over the uncovered portion of the rapidity spectra.  
As can be seen from Figs.~\ref{helium_4pi}~and~\ref{deu_4pi},
cluster multiplicities decrease very fast as collision energy increases.
These results may indicate a decrease of the average nucleon phase space
density which determines the number of $pn$ and $pnp$ combinations for potential
coalescence into $d$ and $^3$He, respectively.

\begin{table}[tbh]
\caption{\label{table:total} Total multiplicity of $^3$He and $d$ in central
Pb+Pb collisions extrapolated to full phase space using two alternative fit functions
(see explanations for the Fits A,B in the text).}
\begin{center}
\begin{ruledtabular}
\begin{tabular}{ccccc}
E$_{beam}$ &
 Fit A & {Fit B} &
 Fit A & Fit B \\
(\agev) & $\langle d \rangle$ & $\langle d \rangle$ & $\langle ^3\text{He} \rangle \cdot 10^{2}$ & $\langle ^3\text{He} \rangle \cdot 10^{2}$\\
\hline
%&\multicolumn{2}{c}{d} &\multicolumn{2}{c}{$^3$He}\\  
20  &$8.42\pm0.43$&$10.46\pm0.54$& $19.92\pm0.72$&$21.74\pm0.79$\\
30 & $5.67\pm0.34$&$7.07\pm0.42$&$11.73\pm0.46$&$17.02\pm0.67$\\
40 &$4.92\pm0.20$&$6.53\pm0.27$&$7.92\pm0.33$&$11.55\pm0.48$\\
80  &$2.74\pm0.17$&$4.60\pm0.28$&$3.46\pm0.19$&$6.03\pm0.33$\\
158  & $1.95\pm0.10$&$3.65\pm0.18$&$1.79\pm0.10$&$3.18\pm0.19$\\
\end{tabular}
\end{ruledtabular}
\end{center} 
\end{table}

In the framework of a statistical thermal model the abundance $N_C$ of
a nucleon cluster of mass $m$, degeneracy factor $g$, charge $q$, and baryon
number $B$ is given by
\begin{equation}\label{eq_therm}
N_C=\frac{gV}{\pi^2}{m^2}{T}{K_2(m/T)}\exp{\left(\frac{B\mu_B+q\mu_q}{T}\right)}~~,
\end{equation}
where $V$, $T$, $\mu_B$, $\mu_q$, and $K_2$ are the source volume, temperature,
baryochemical potential, charge potential, and Bessel function of the second kind.
Such models have been able to reproduce the multiplicities of different types
of particles in elementary and heavy-ion interactions. There are several
parameterizations for the thermal fireball parameters $T$, $\mu_B$, and $V$ 
(or equivalently the fireball radius $R$)
over a wide range of nuclear collision energies from AGS to
LHC~\cite{pbm,cleym,shm,vbg}. The overall average of these predictions
at SPS energies is given in Table~\ref{table_param}. The listed uncertainty is
taken as half of the difference between the highest and lowest value for the
fireball parameters provided by the various parameterizations. The $\mu_q/T$ values
were obtained from NA49 measurements of the $\pi^+$/$\pi^-$ ratio~\cite{na49_pika} as
\begin{equation}
\frac{\mu_q}{T}=\frac{1}{2}\ln{\left(\frac{\pi^+}{\pi^-}\right)}
\end{equation}

\begin{table*}[tb]%[H] add [H] placement to break table across pages
\begin{center}
\begin{ruledtabular}
\caption{\label{table_param}The fireball thermodynamical parameters (the average
of parameterizations given in~\cite{pbm,cleym,shm,vbg}) used in thermal model
calculations, and the predicted total multiplicities of $d$ and $^3$He in central Pb+Pb collisions.}
\begin{tabular}{ccccccc}
E$_{beam}$ & $T$  & $\mu_B$ & $R$ & $\mu_q/T$ & $\langle N_{d} \rangle$  & $\langle N_{He}\rangle \cdot10$ \\
(\agev) & (MeV) & (MeV) & (fm) & & & \\
\hline
20 & $133 \pm 2$ & $472 \pm 8$ & $8.2 \pm0.2$ & $-0.075 \pm 0.008$ & $6.56 \pm 0.78$ &$2.28 \pm 0.40$  \\
30 & $140 \pm 2$ & $417 \pm 7$ & $8.3 \pm 0.1$ & $-0.064 \pm 0.006$ &$5.10\pm 0.51$ &$1.46 \pm 0.22$ \\
40 & $145 \pm 2$ & $377 \pm 8$ & $8.6 \pm 0.1$ &$-0.053 \pm 0.005$ &$4.41 \pm 0.49$ & $1.09 \pm 0.18$\\
80 & $153 \pm 3$  & $294 \pm 9$ & $9.3 \pm 0.2$ &$-0.047 \pm 0.005$ & $3.11 \pm 0.41$ & $0.54 \pm 0.11$ \\
158 & $158 \pm 4$ & $224 \pm 10$ &$10.1 \pm 0.7$ &$-0.036 \pm 0.004$ & $2.00 \pm 0.28$&$0.26 \pm0.05$ \\
\end{tabular}
\end{ruledtabular}
\end{center}
\end{table*}

Using these fireball parameters the mean multiplicities of
$d$ and $^3$He were computed at all five collision energies according to Eq.~\ref{eq_therm}.
The results are plotted in Figs.~\ref{helium_4pi}~and~\ref{deu_4pi} with blue circles 
and are listed in the last two columns of Table~\ref{table_param}.
The overall uncertainties of the total yields were estimated by standard error propagation.
  
As can be seen, thermal model calculations are capable of reproducing
the energy dependence of the cluster multiplicities not only qualitatively
but also quantitatively.
The deviation of the calculations from the measured abundances
does not exceed 2 standard deviations (see insets in Figs.~\ref{helium_4pi}~and~\ref{deu_4pi}).
It seems that there might be a systematic underprediction for the yield of $d$, which
however cannot be claimed to be significant due to the correlated systematic uncertainty
of the extrapolation to full phase space.
%It appears at first glance that such a good agreement between experimental
%data on cluster production and thermodynamical model calculations looks
%surprising
%since one expects that chemical equilibrium of compound nucleon systems
%with the environment reaches later in time than the chemical freezeout in
%the fireball has happened. The more diluted and colder matter is needed
%to make a net (fusion minus decomposition) rate positive for bound nucleon
%systems of several MeV binding energy.
%It has been recognized, however, that the light
%nuclei production process is basically determined by the entropy per baryon
%of the medium that is fixed at the chemical freezeout~\cite{entr1,entr2}.
%During further isentropic expansion stage a dynamic equilibrium has reached
%for the light nuclei population: the loss due to disintegration are balanced by
%the gain in newly produced clusters. 

\begin{figure}
\includegraphics[width=1.0\linewidth]{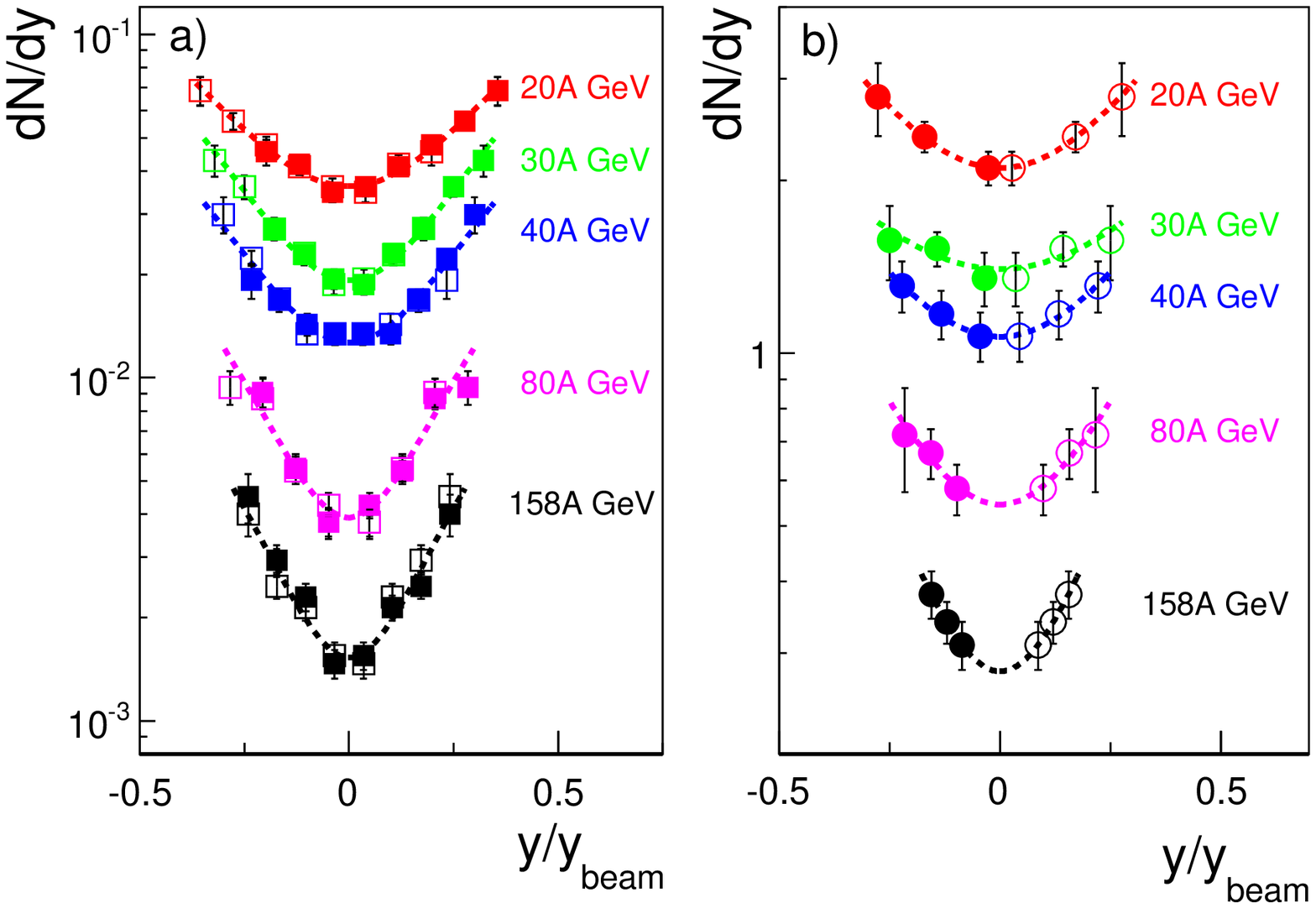}
\caption{(Color online) Rapidity distributions for $^3$He (a) and
deuterons (b) from central Pb+Pb collisions at 20$A$-158\agev~versus normalized
rapidity $y/y_{beam}$.
The solid symbols show the measurements and the open symbols
represent the data points reflected around mid-rapidity.
The error bars correspond to the quadratic sum of statistical and systematic
errors. Dashed lines indicate parabolic fits to the rapidity spectra
(see text for more detail).}
\label{dndy_dhel}
\end{figure}

\begin{figure}
\includegraphics[width=0.95\linewidth]{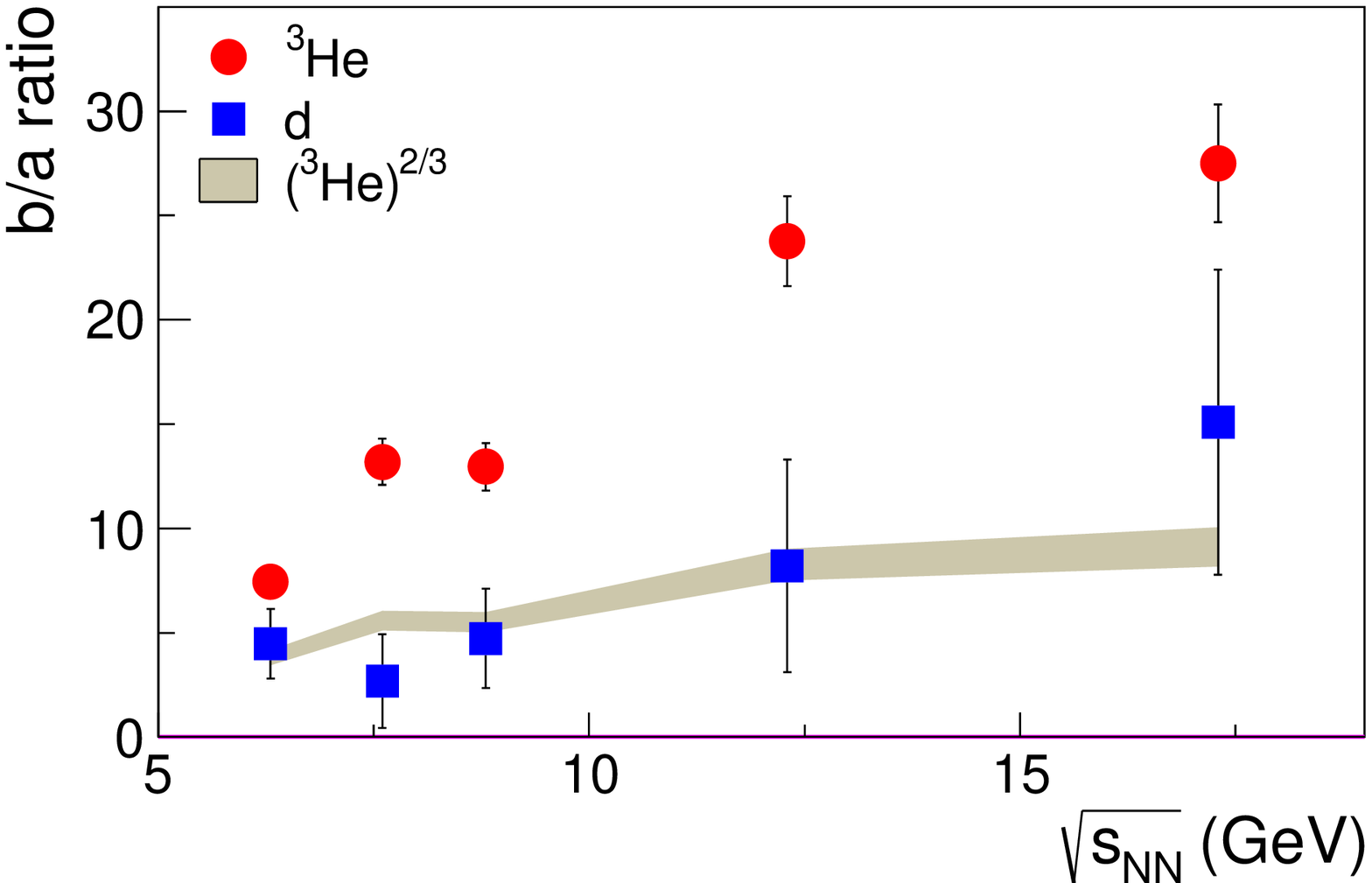}
\caption{(Color online) The ratio $b/a$ of the fit parameters obtained for the rapidity spectra
of $^3$He (dots) and $d$ (squares) from central Pb+Pb collisions as a function of $\sqrt{s_{NN}}$.}
\label{concavity}
\end{figure}

To inspect how the shape of the rapidity spectra for light nuclei
varies with collision energy and atomic mass number, cluster yields
are plotted in Fig.~\ref{dndy_dhel} as a function
of the normalized rapidity $y/y_{beam}$.  All the rapidity distributions
are concave. In order to quantify the changes the data were fitted
with a parabola $a+b \, (y/y_{beam})^2$ (the fits are shown by
dashed lines). The ratio of the fit parameters $b/a$ ({\it relative concavity})
for $^3$He and $d$ is plotted in Fig.~\ref{concavity} as a function of
$\sqrt{s_{NN}}$. One observes that the relative concavity of the rapidity
distributions for light nuclei tends to increase with increasing beam
momentum and cluster mass. Such a behavior of the invariant yields versus
rapidity was earlier observed at AGS energies~\cite{clust_e864},
where the relative concavity of the yield of clusters with atomic mass number
from $A=2$ to 4 progressively increases with $A$ in the range of transverse
momentum $0.1 < p_t/A < 0.2$~\gevc.

Assuming that coalescence is the dominant process
of cluster formation close to mid-rapidity, one expects the relative concavity
of the rapidity spectra for $^3$He to increase by the power of 3/2 relative
to that for $d$ (2/3 in case of the reverse order).
In Fig.~\ref{concavity}, the shaded area shows the $b/a$ ratio for
the $^3$He spectra to the power of 2/3. One indeed observes that
the measured shapes are consistent with this expectation.    

\subsection{The mass number dependence of light nuclei production}

\label{penalty_fact}
\begin{figure}
\includegraphics[width=.95\linewidth]{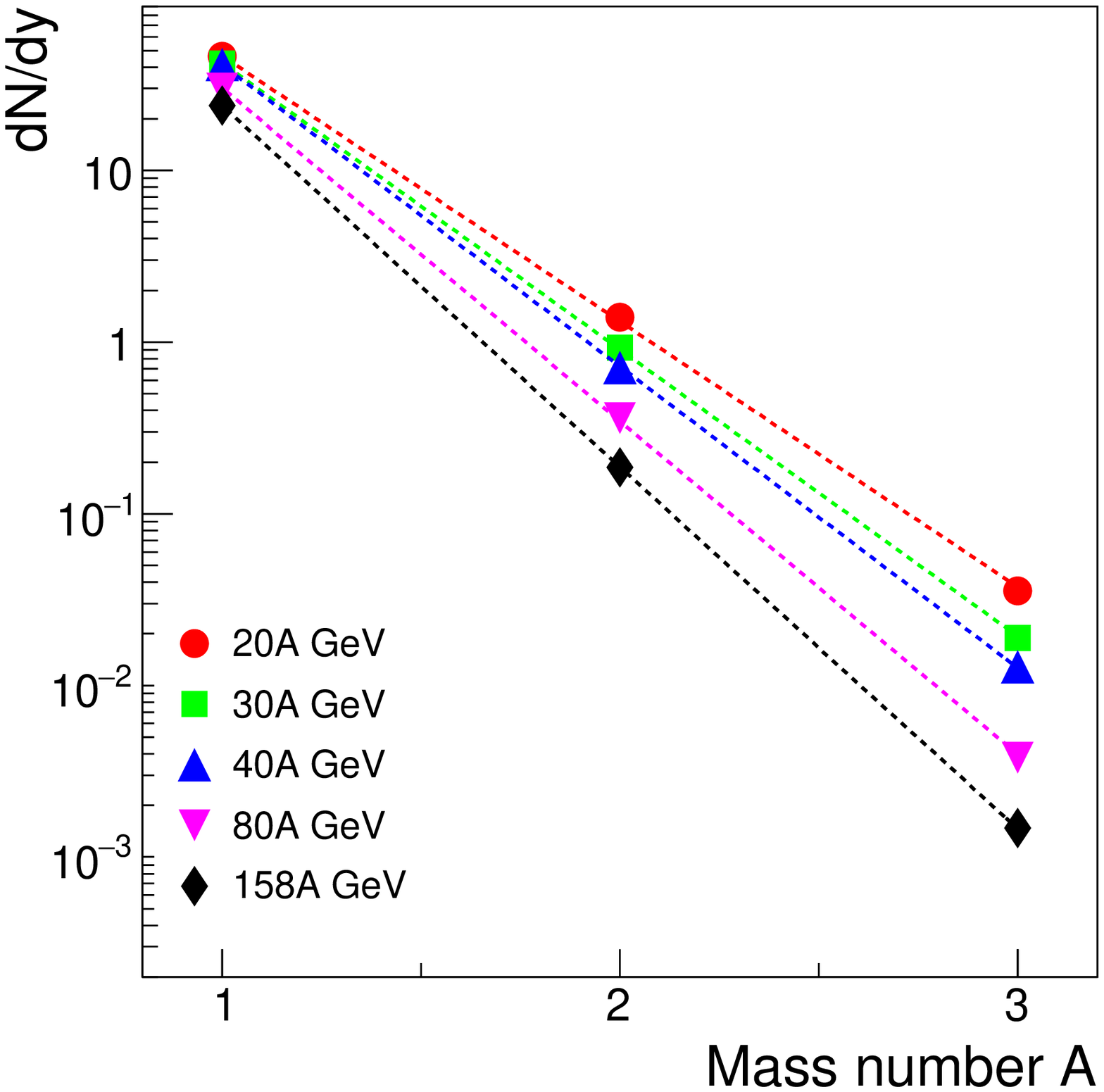}
\caption{(Color online) Values of \dndy~at mid-rapidity (see text) as a function of 
mass number $A$ ($p, d$, $^3$He) from central Pb+Pb collisions at 20$A$-158\agev.
The dashed lines represent the fit to an exponential dependence (see text Eq.~\ref{penalty}). 
Values of \dndy~for $d$ were divided by the spin factor 3/2 (see text for more detail).}
\label{penal1}
\end{figure}

\begin{figure}
\includegraphics[width=1.0\linewidth]{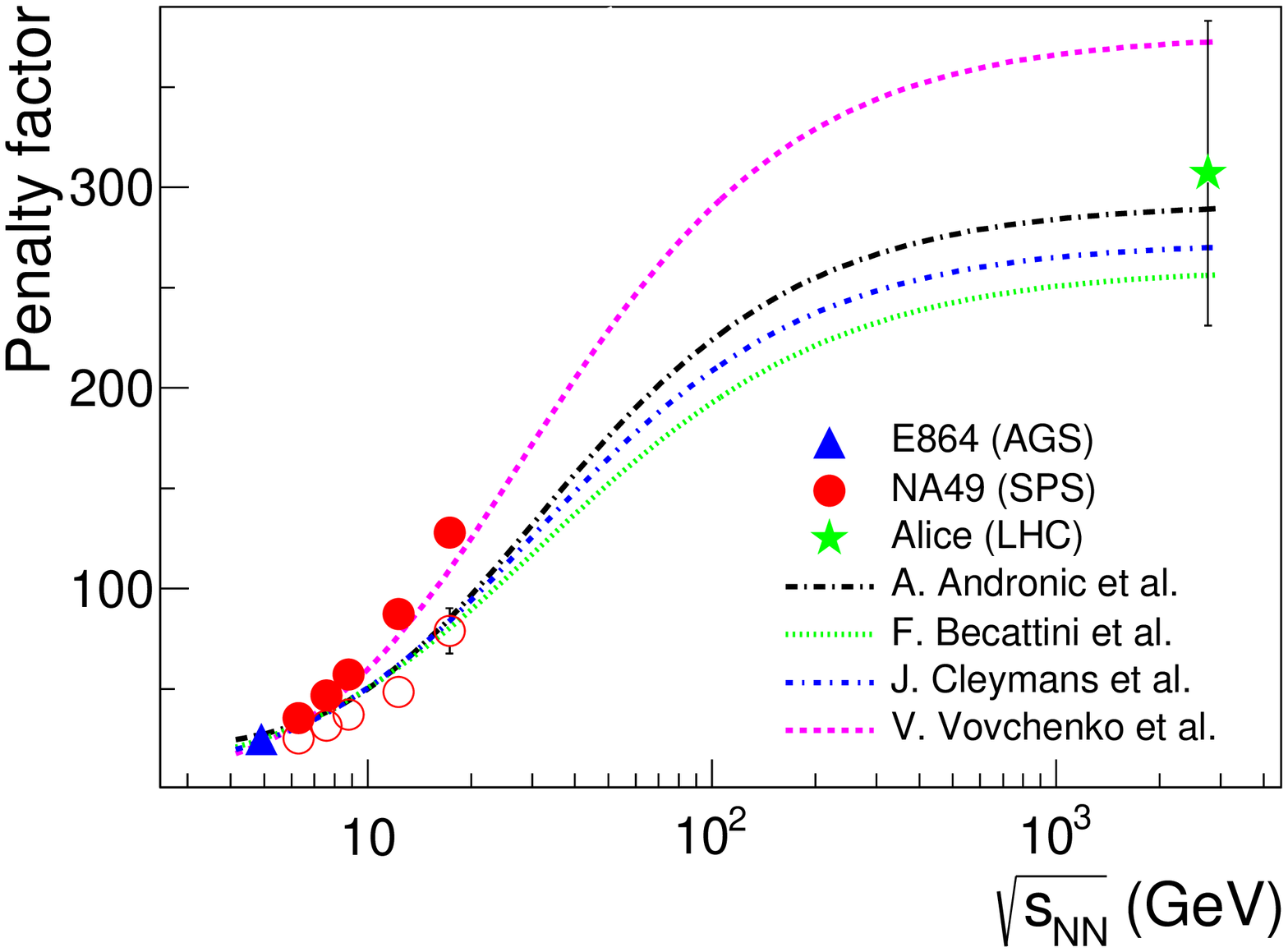}
\caption{(Color online) Penalty factor from the cluster yields at mid-rapidity
(closed symbols) and 4$\pi$ multiplicities (open circles) in central A+A
collisions. Red circles represent the NA49 data, the AGS measurement
(blue triangle) is from~\cite{clust_e864}, and the green star indicates
the result from the ALICE experiment~\cite{alice}. The thermal statistical
model estimates are from~\cite{pbm,cleym,shm,vbg} (see text for detail).}
\label{penal2}
\end{figure}

Typically cluster production yields change drastically with the atomic mass number
$A$ and can be cha\-rac\-te\-ri\-zed by a parameter $P$, the {\it 'penalty factor'} for
adding an extra nucleon to the system. Figure~\ref{penal1} presents the
$A$ dependence of the mid-rapidity yield \dndy~for $p$, $d$, and $^3$He. The NA49
measurements for protons are taken from Refs.~\cite{na49_ppbar, na49_ppbar_milica}.
The data points for $d$ and $^3$He are the numerical values of the parabolic
fits to the rapidity spectra in Figs.~\ref{dndy_hel}~and~\ref{dndy_deu} at mid-rapidity ($y=0$).
In a statistical approach the particle production rate is proportional to its spin
degeneracy factor (2J+1), so it is reasonable to divide the deuteron rates by the factor 3/2
% added 25.05.16 V.I. Kolesnikov
as it was done in Ref.~\cite{e864_prl}.
% end of adding.
The penalty factor $P$ was then obtained from a fit to the atomic mass number dependence of \dndy~ 
at mid-rapidity with an exponential function of the form:
\begin{equation}\label{penalty}
const/P^{A-1}  .
\end{equation}
The fit results are drawn in Fig.~\ref{penal1} as dashed lines
and the fitted values of the parameter $P$ are shown in Fig.~\ref{penal2} as a function of
$\sqrt{s_{NN}}$.  

The same analysis can be performed on the total yields of nucleon clusters.
For protons the rapidity spectra at 40$A$ and 158\agev~
from~\cite{na49_ppbar_milica} were extrapolated to the full phase space
employing a parametrization by the sum of three Gaussian distributions
(as described above). At other energies, however, proton measurements
over a phase space region sufficient for extrapolation to 4$\pi$ are not
available. Thus at these energies the penalty factors and their uncertainties 
had to be calculated from the integrated yields for $^3$He and $d$ only.  
The results are plotted in
Fig.~\ref{penal2} as open symbols supplementing the data points obtained
at AGS, SPS and LHC energies derived from \dndy~at mid-rapidity.
The measurement at the lowest energy is from the experiment
E864~\cite{clust_e864} (10\% most central Au+Pb collisions at $\sqrt{s_{NN}}=4.6$~GeV),
the LHC results are from the ALICE Collaboration~\cite{alice}
(20\% most central Pb+Pb collisions at $\sqrt{s_{NN}}=2.76$~TeV).

The excitation function for the penalty factor rises rapidly at small
collision energies (the slope for the points based on mid-rapidity \dndy~is greater than
for the 4$\pi$ multiplicity data) and appears to level off
at higher energies. Such a saturation behavior can be explained within thermal
statistical models where the penalty factor $P$ for cluster yields is determined by the
Boltzmann factor exp$\left[(m-\mu_{B})/T\right]$~(see e.g. Ref.~\cite{therm5}),
with $\mu_{B}$, $T$, and $m$ being the baryochemical potential, freezeout temperature,
and nucleon mass, respectively.
Employing the parameterizations for the energy dependence of $T$ and
$\mu_{B}$ established in Refs.~\cite{pbm,shm,cleym,vbg}, 
the Boltzmann factor was computed over the region of collision energies from
$\sqrt{s_{NN}}=4$~GeV to 3 TeV.  The calculated excitation
functions are drawn in Fig.~\ref{penal2} with lines of different types.
As can be seen, thermal model predictions are in qualitative agreement
with the measured penalty factors.
   
\begin{figure*}
\begin{minipage}[b]{0.95\linewidth}
\includegraphics[width=0.95\linewidth]{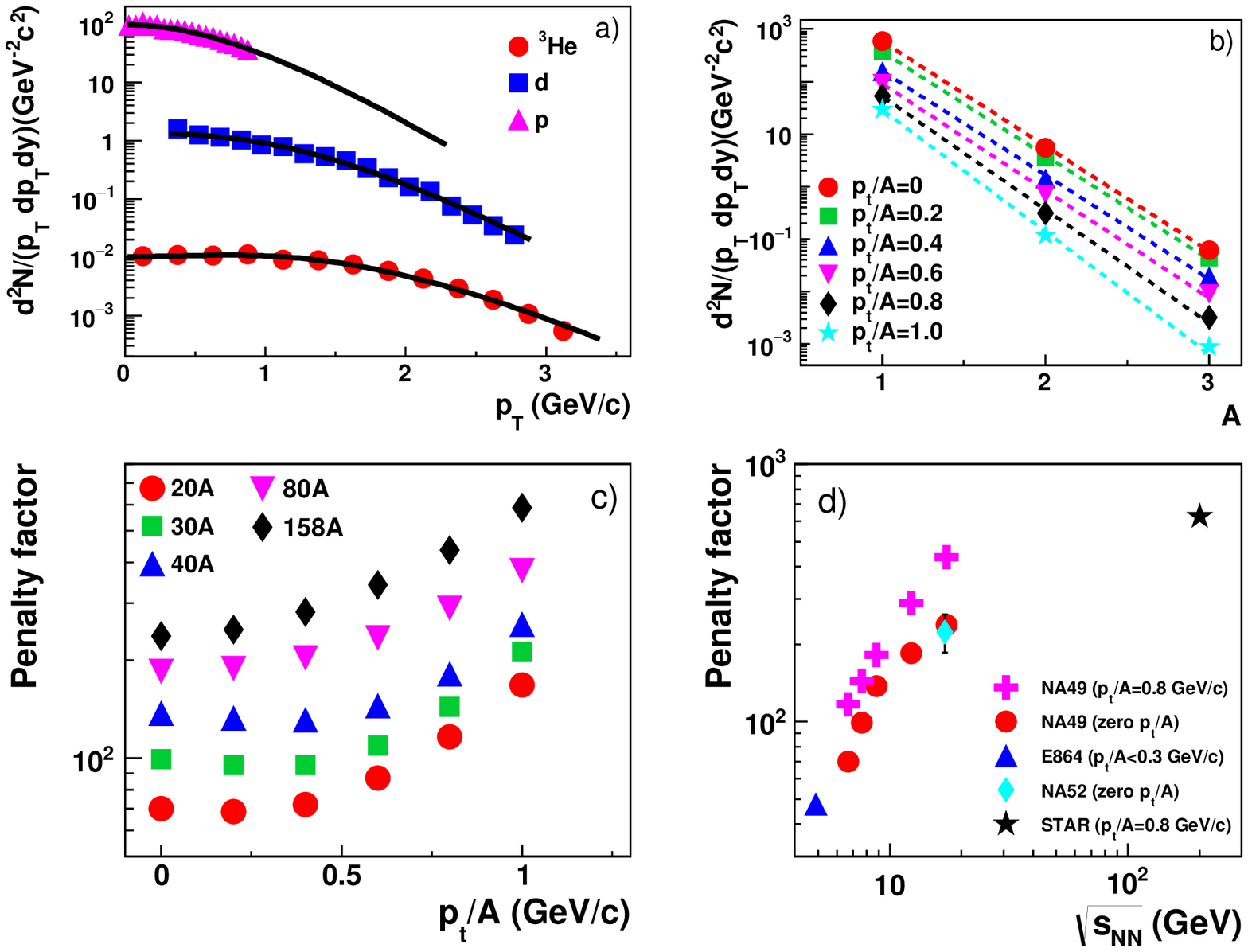}
\end{minipage}
\caption{(Color online) a) Invariant mid-rapidity $p_t$ spectra of $p$, $d$
and $^3$He from central Pb+Pb collisions at 30\agev; b) Invariant yield of clusters
at several values of $p_t/A$ from central Pb+Pb collisions at 30\agev~
(values for $d$ were divided by the spin factor 3/2); c) Penalty factor $P$ from
the cluster yields at several values of $p_t/A$ from central Pb+Pb collisions at 20$A$-158\agev;
d) Excitation function for the penalty factor at $p_t/A$ near zero
and $p_t/A=0.8$~\gevc. AGS data are from~\cite{clust_e864}, the result from
the STAR Collaboration was reported in~\cite{star_pen}.}
\label{penalty_pt}
\end{figure*}

For a complete picture, one should bear in mind that there
exist more data on the penalty factor for nucleon clusters detected in more restricted 
phase space regions. For example, analyzing the yields of light nuclei
at $p_t/A<300$~\mevc, a penalty factor of $48\pm3$ was found
by the E864 experiment in the 10\% most central Au+Pb interactions at beam
energy of 11.5\agev~\cite{e864_prl}. A value of about $223\pm38$
was obtained near zero $p_t$ by the NA52 experiment from minimum bias
Pb+Pb collisions at 158\agev~\cite{na52_penal}. Reference~\cite{star_pen} 
reports a penalty factor of about 625 (at $p_t\sim0.8$~\gevc~per
nucleon) that was deduced using measurements by the STAR experiment
in the 12\% most central Au+Au collisions at $\sqrt{s_{NN}}=200$~GeV~\cite{star_penalty}.
All the above mentioned values of $P$ were obtained for small regions
of the final state phase space, thus they cannot be directly compared
to those extracted from the integrated data. This is simply due to the
%because of the fact that
strong radial flow of baryons in heavy-ion collisions resulting
in a non-trivial pattern of space-momentum correlations at freezeout. This affects
the probability of cluster formation differently in different phase space cells.
In order to illustrate this, light nuclei production yields were studied more 
differentially in bins of $p_t/A$. The results are presented
in Fig.~\ref{penalty_pt}.  As an example, panel a) shows invariant $p_t$ spectra
of $p$, $d$ and $^3$He near mid-rapidity from central Pb+Pb collisions
at 30\agev~(data for protons were taken from Ref.~\cite{na49_ppbar}).
The distributions were fitted to a sum of two exponentials
(shown by lines in Fig.~\ref{penalty_pt})) 
with proper error estimates assigned to each data point of the
$p_t$ spectrum. The invariant yields (with uncertainties) for $p$, $d$, and
$^3$He were calculated from the integrals of the fit functions at several values of $p_t$ per nucleon
in the range $p_t/A$  from 0 to 1.0~\gevc~. Each triple is shown in Fig.~\ref{penalty_pt}(b)
and was fitted to Eq.~\ref{penalty}. The resulting values of the penalty factor
$P$ for each energy are plotted in Fig.~\ref{penalty_pt}(c) as a function
of $p_t/A$. An in\-te\-res\-ting regularity is observed for the $p_t$ dependence of the penalty
factor: 1) the dependence on $p_t/A$ is similar for all energies, but the magnitude increases
with increasing collision energy;
2) the penalties vary slowly at low $p_t/A$ and  begin to rise faster above
$p_t/A\sim0.5$~\gevc.
The excitation functions of the penalty factor for zero
$p_t$ and $p_t/A=0.8$~\gevc~ are plotted in  Fig.~\ref{penalty_pt}(d).
As can be seen, the measurements at AGS and RHIC energies are
consistent with the trend shown by the data from the SPS.

In the framework of both the thermal and coalescence approaches the penalty
factor is related to the average phase space density of single nucleons
$\langle f_{N}(x,p)\rangle$. From a microscopic point of view, $\langle f_{N}\rangle$
results from an interplay of the stopping power and the strength of 
flow in the reaction. As collision energy increases, nucleon stopping becomes
weaker, while the collective transverse motion gets stronger, thus explaining
the observed trend of the penalty factor to increase with $\sqrt{s_{NN}}$.     

It was also found experimentally that the ratio of deuterons to protons
at mid-rapidity is nearly constant over the whole collision centrality range in Pb+Pb
in\-te\-rac\-tions at the top SPS energy~\cite{deu2_na49}. This finding (taking into account
that the $d/p$ ratio can be related to $\langle f_{N}\rangle$) implies a small
variation of the average baryon phase space density with collision centrality and thus
offers an explanation for the good agreement between the NA49
measurement of the penalty factor near zero $p_t$ from central Pb+Pb collisions
at 158\agev~and the one obtained by the NA52 Collaboration from
a minimum bias data set (see Fig.~\ref{penalty_pt}(d)).  

%Study the bulk thermodynamical  properties of nuclear matter
%is also possible by means of transversal distributions of specie.
%So, we go to the next section.
\begin{figure*}
\begin{minipage}[b]{0.95\linewidth}
\includegraphics[width=0.95\linewidth]{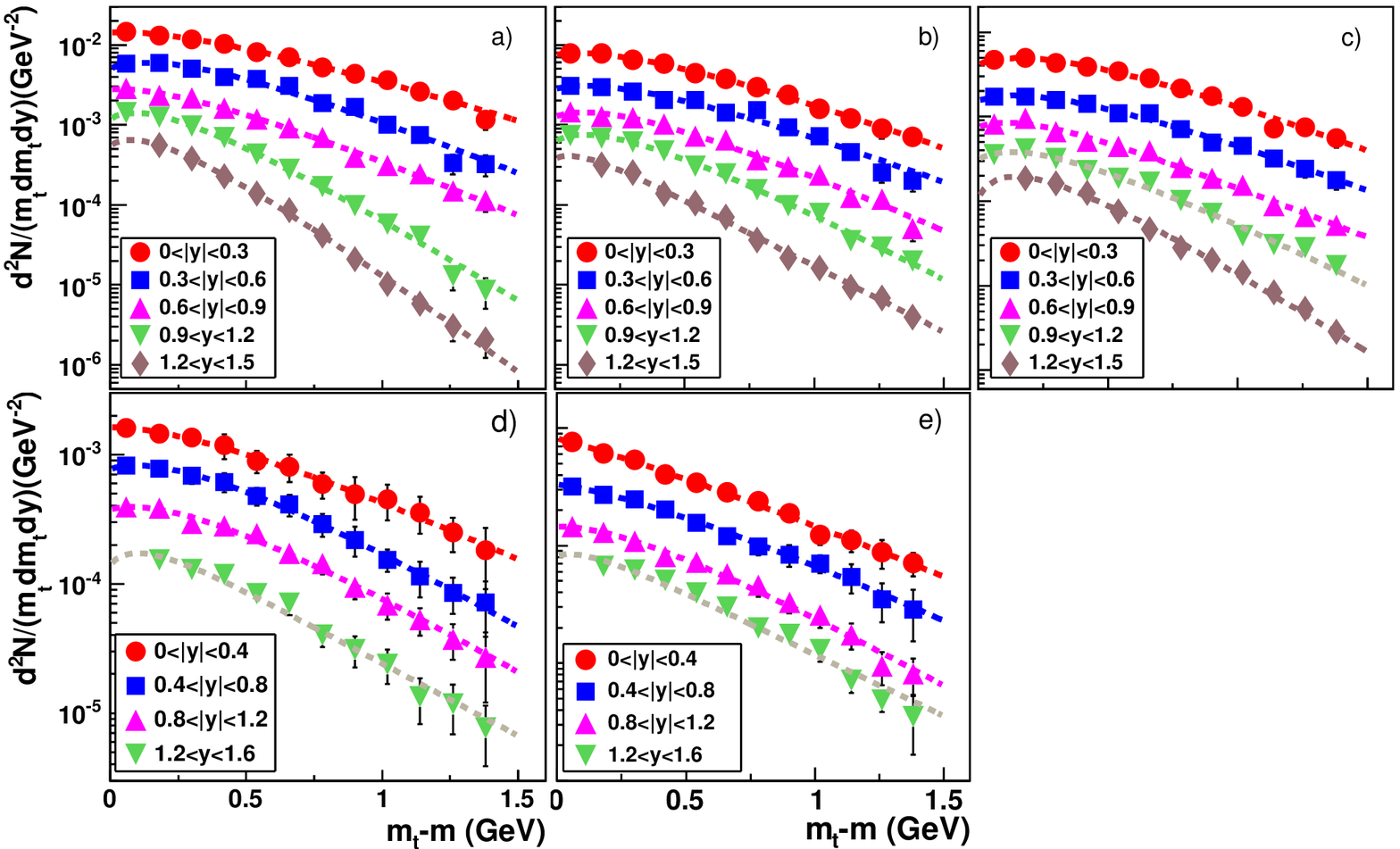}
\end{minipage}
\caption{(Color online) $m_t$ spectra of $^3$He in rapidity slices from
central Pb+Pb collisions at 20$A$ (a), 30$A$ (b), 40$A$ (c), 80$A$ (d), and 158\agev~(d).
The spectra are scaled down by a factor of 3 successively. Only statistical
errors are plotted. Dashed curves indicate the fits to the spectra with
a sum of two exponential functions.}
\label{all_mt_hel}
\end{figure*}  

\subsection{Analysis of transverse mass spectra}

This section discusses systematic dependences of transverse mass spectra of clusters 
on the collision energy, rapidity, and particle mass.
Commonly, $m_t$ distributions are examined either individually
in terms of the characteristic inverse slope parameter $T_{eff}$
({\it effective temperature}) or simultaneously in the framework
of a blast-wave (BW) model. 

\begin{figure}
\includegraphics[width=0.95\linewidth]{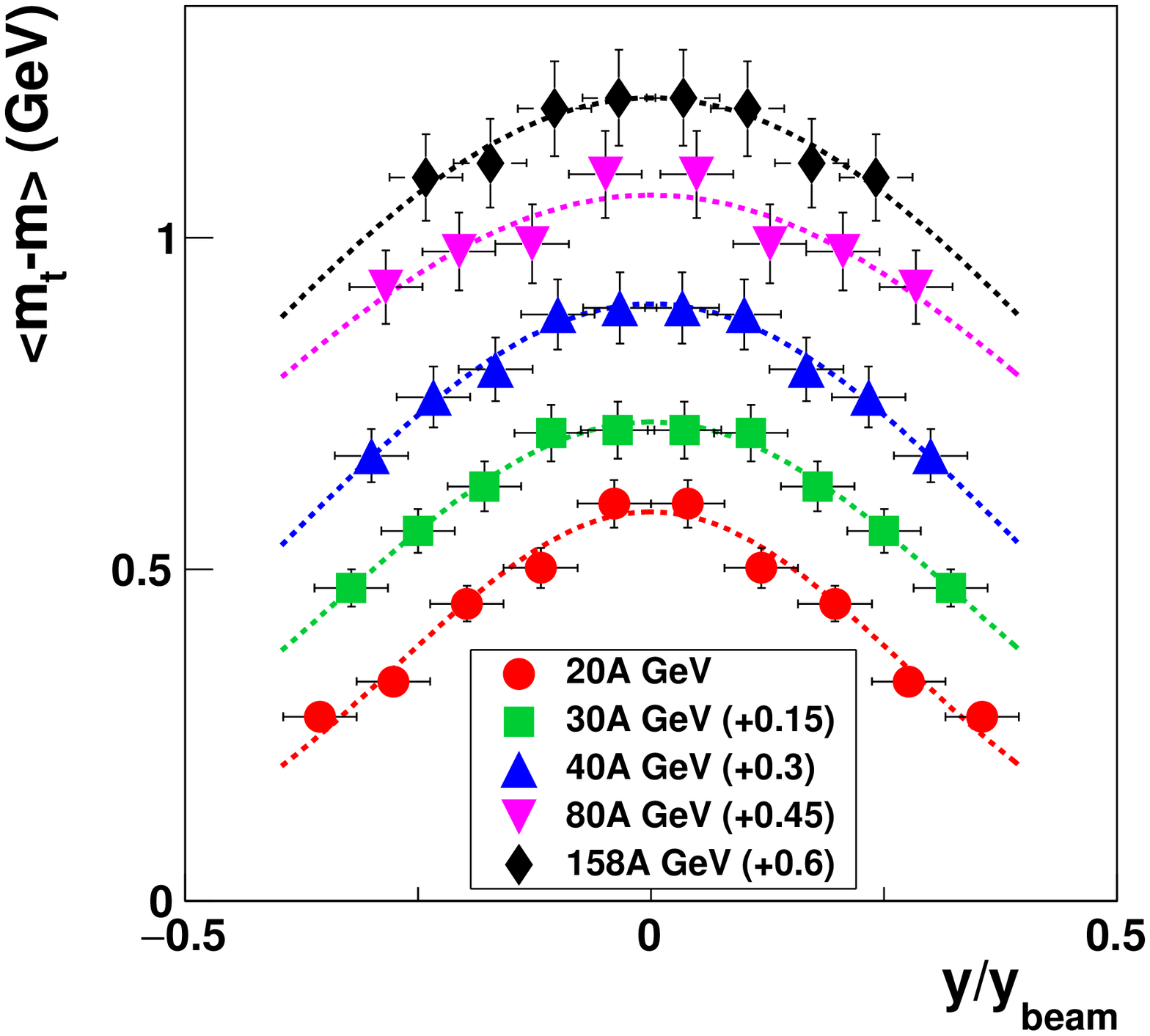}
\caption{(Color online) Mean transverse mass for $^3$He as a function
of $y/y_{beam}$ from central Pb+Pb at 20$A$-158\agev. Results for the forward
and backward hemispheres were averaged (see text). Dashed lines indicate fits
to a Gaussian.}
\label{meanmt_hel}
\end{figure}

\begin{figure}
\begin{center}
\begin{minipage}{0.95\linewidth}
\includegraphics[width=0.95\linewidth]{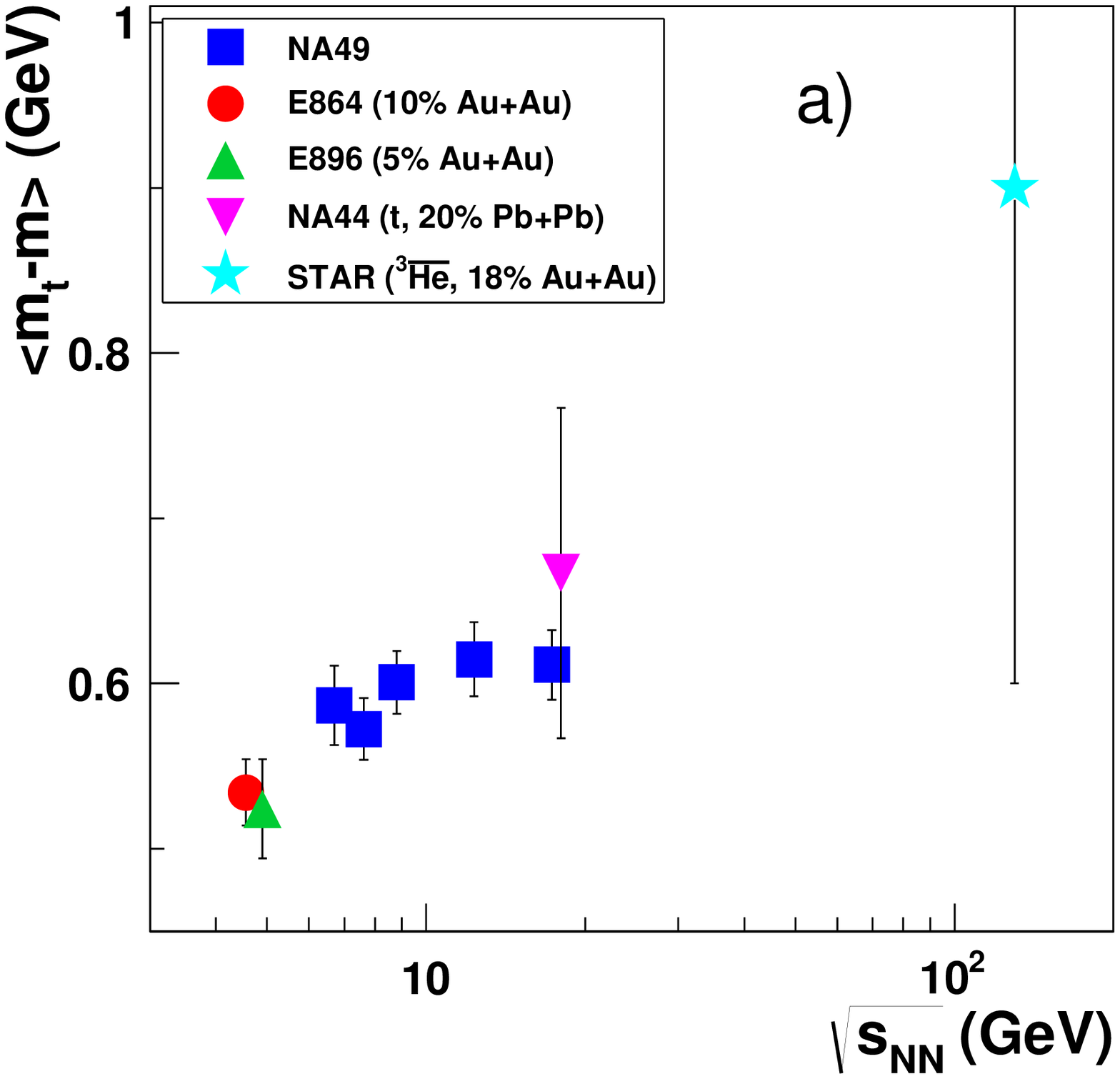}
\end{minipage}
\begin{minipage}{0.95\linewidth}
\includegraphics[width=0.95\linewidth]{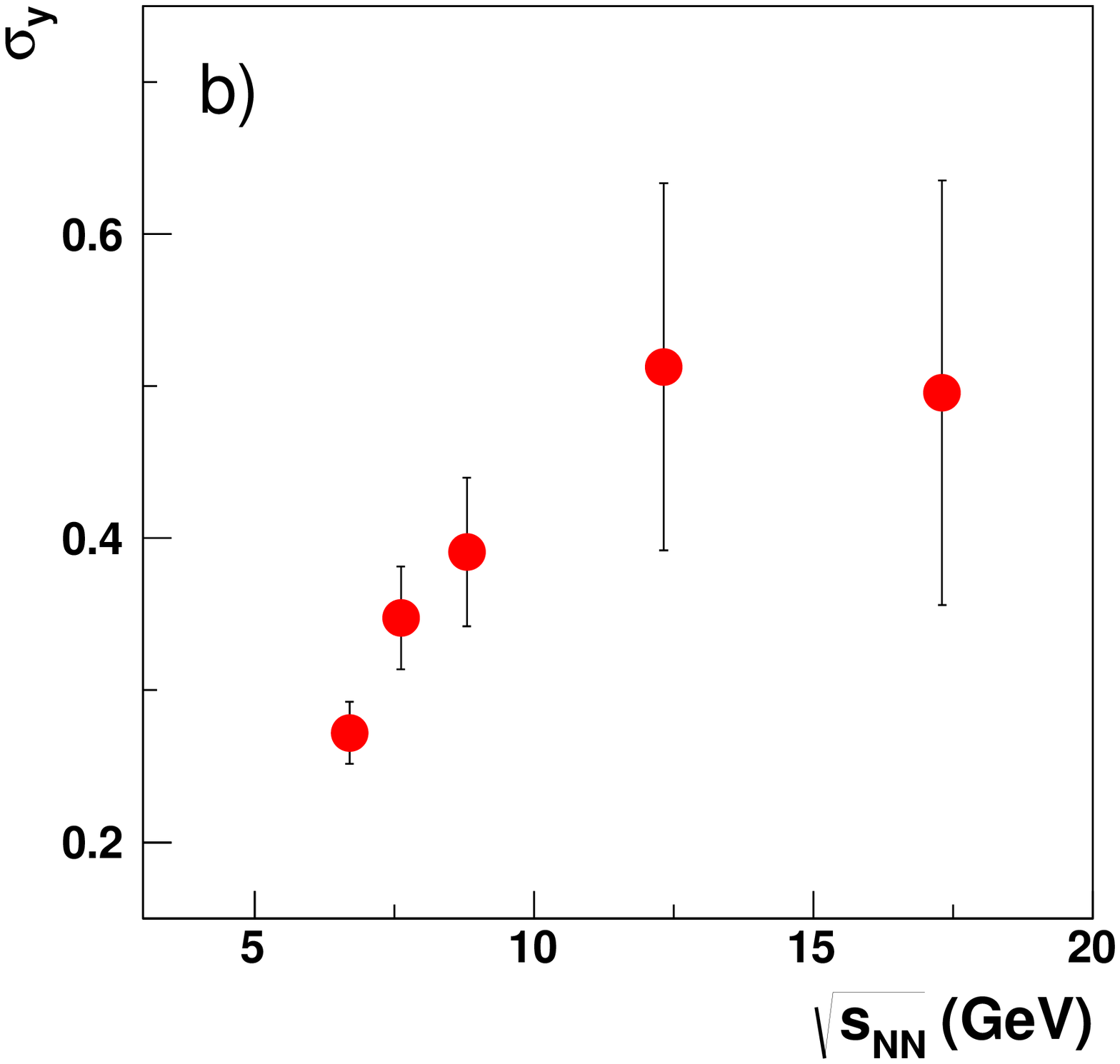}
\end{minipage}
\caption{\label{meanmt_energy}(Color online) a) $\langle m_t\rangle -m$ at mid-rapidity
vs. $\sqrt{s_{NN}}$ for $A=3$ clusters from central A+A collisions at SPS,
AGS, and RHIC energies. b) The width $\sigma_{y/y_{beam}}$ of the Gaussian fitted 
to the rapidity dependence of $\langle m_t\rangle -m$ 
(see Fig.~\ref{meanmt_hel}) for $^3$He versus $\sqrt{s_{NN}}$.}
\end{center}
\end{figure}

The first method was applied to extract
both $T_{eff}$ and the mean transverse kinetic energy
$\langle m_t\rangle -m$. Figure~\ref{all_mt_hel} shows the fully corrected $m_t$ spectra of
$^3$He nuclei in rapidity slices at five bombarding energies.
The rapidity binning, indicated in the figure, was the same as
in the previous section and data from the bins symmetric re\-la\-ti\-ve
to mid-rapidity $y=0$ were combined in order to decrease statistical fluctuations.
The average transverse energy $\langle m_t\rangle -m$ was deduced from the measured data points
combined with the integral over the unmeasured $m_t$-range. The latter was computed
by fitting the spectra with a sum of two exponential functions.
The results for $\langle m_t\rangle -m$ of $^3$He are presented in
Fig.~\ref{meanmt_hel} as a function of normalized rapidity $y/y_{beam}$.
As can be seen, the rapidity dependence at all energies follows a bell-like shape. Thus,
Gaussian fits were applied keeping the position of the maximum fixed at mid-rapidity $y=0$.
The two other parameters of the Gaussians ($\langle m_t\rangle -m$ at mid-rapidity and width $\sigma_y$))
are plotted in Fig.~\ref{meanmt_energy} as a function of center-of-mass
energy $\sqrt{s_{NN}}$.
The drawn uncertainties are the fit errors. The overall systematic
uncertainty for $\langle m_t\rangle -m$ at mid-rapidity was estimated to amount
to less than 5\%. As can be seen from Fig.~\ref{meanmt_energy}(a), where
the present results are shown along with the measurements at AGS~\cite{clust_e864,clust_e896}
and RHIC~\cite{dbar_star} energies, the mean transverse mass for $^3$He as a function of the
collision energy qualitatively follows the trend observed for hadrons in
central A+A collisions~\cite{na49_pika}: rising at low energies and
leveling off in the SPS energy region. The shape of the transverse mass spectra
is mainly determined by the parameters characterizing the source (temperature,
pressure and collective velocity profile). Consequently,  the results suggest
that at SPS energies the variations in these basic fireball parameters are small.

\begin{figure}
\includegraphics[width=1.0\linewidth]{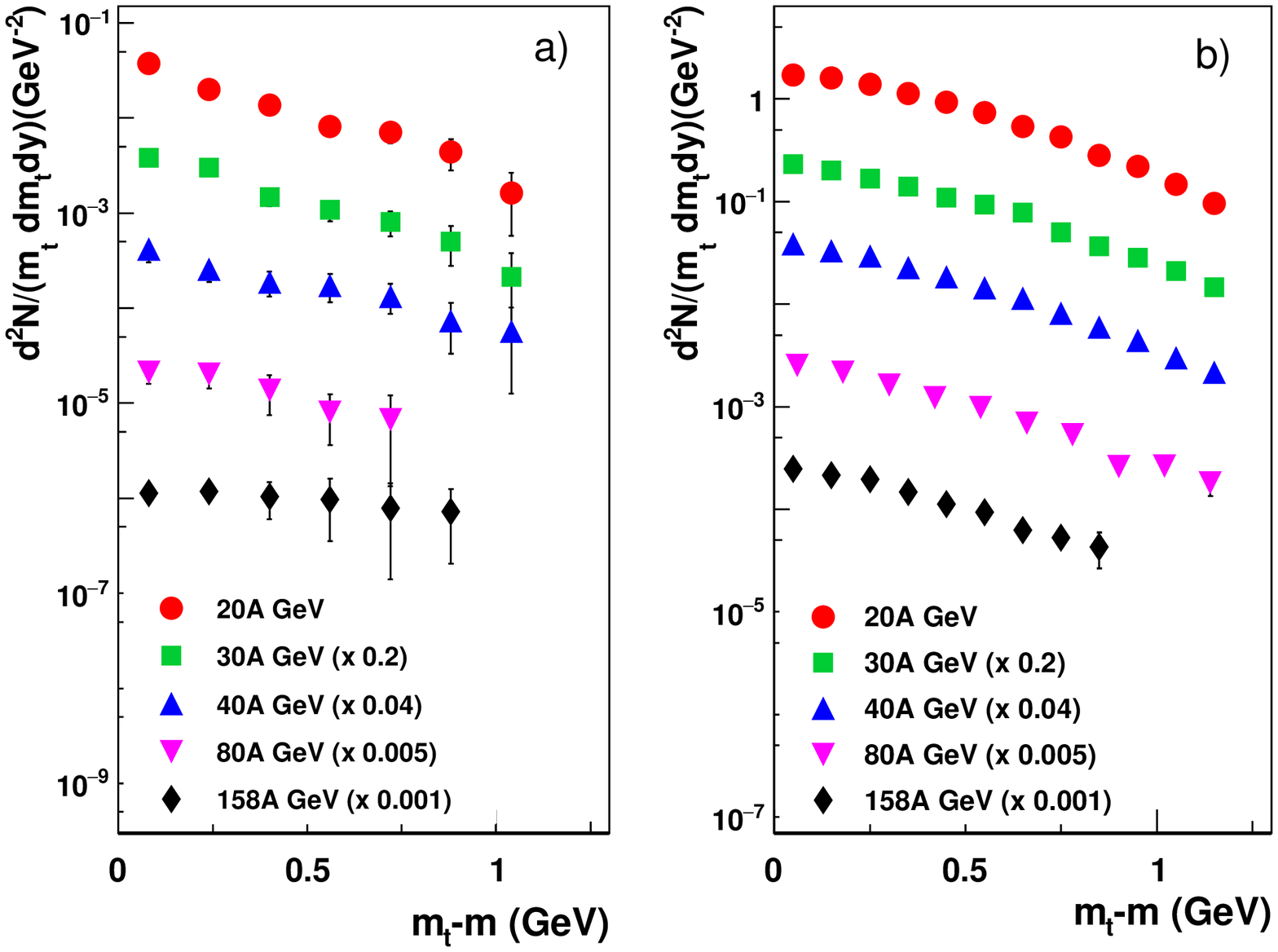}
\caption{(Color online) Invariant $m_t$ spectra from central Pb+Pb collisions
at 20$A$-158\agev~for tritons (a) and deuterons (b).}
\label{all_mt_dt}
\end{figure}
For $d$ and $t$ the acceptance (defined by requiring TOF information) is more
restricted in rapidity and transverse mass (see Fig.~\ref{acc_all}). 
Therefore, in order to get coverage in $m_t$$-m$
sufficient for examination of the spectra shapes,
the rapidity interval was enlarged to $\Delta y\approx0.6$ and $\Delta y>1.0$
in case of $d$ and $t$, respectively.
Figure~\ref{all_mt_dt} presents the fully corrected transverse mass spectra for
$d$ and $t$ in these rapidity intervals from central Pb+Pb collisions at five bombarding
energies. 

The numerical values of $\langle m_t\rangle -m$ for $d$ and $^3$He at
mid-rapidity are given in Table~\ref{table2}. For $^3$He the numbers for
$\langle m_t\rangle -m$ along with their uncertainties are taken from the
Gaussian fits in Fig.~\ref{meanmt_hel}. The values for deuterons
represent extrapolations to mid-rapidity from the measured rapidity range
(the average rapidity $\langle y \rangle$ is -0.3, -0.35, -0.4, -0.7,
and -0.7 at 20$A$, 30$A$, 40$A$, 80$A$, and 158\agev, respectively) under the explicit
assumption of a Gaussian rapidity dependence of $\langle m_t\rangle -m$
with the width parameter $\sigma_y$ taken from the results for $^3$He.

Since transverse mass spectra of clusters are not well described by an exponential
function (see Fig.~\ref{all_mt_hel})
they cannot be characterized by a single slope parameter in most cases. The estimates for
$T_{eff}$ were therefore obtained by fitting the spectra with exponential functions
excluding the region of $m_t-m<0.5$~GeV. The results are listed in Table~\ref{table2}. 
 
\begin{table}[tbh]
\caption{\label{table2} The mean transverse kinetic energy and effective slope
parameter for $d$ and $^3$He at mid-rapidity.}
\begin{center}
\begin{ruledtabular}
\begin{tabular}{ccccccc}
E$_{beam}$
&
$\langle m_t\rangle -m$ & $T_{eff}$ &
$\langle m_t\rangle -m$ & $T_{eff}$ \\
(\agev) & (MeV) & (MeV) & (MeV) & (MeV)\\
\hline
&\multicolumn{2}{c}{$d$} & \multicolumn{2}{c}{$^3$He}\\ 
20  &~~$463\pm28$~~&~~$317\pm18$~~&~~$581\pm29$~~&~~$406\pm20$~~\\
30 & ~~$468\pm28$~~&~~$320\pm20$~~&~~$573\pm30$~~&~~$424\pm22$~~\\
40 &~~$453\pm27$~~&~~$328\pm21$~~&~~$600\pm35$~~&~~$425\pm25$~~\\
80  &~~$476\pm28$~~&~~$368\pm41$~~&~~$612\pm44$~~&~~$525\pm60$~~\\
158  & ~~$517\pm38$~~&~~$390\pm55$~~&~~$610\pm46$~~&~~$512\pm50$~~\\
\end{tabular}
\end{ruledtabular}
\end{center} 
\end{table}

For the case of tritons, however, extraction of the slope parameter of
the spectra becomes problematic. At low $m_t$ the yields are measured far
away from the central rapidity, while at larger $m_t$ the acceptance for tritons is near
mid-rapidity. Therefore the shape of the triton spectra is strongly modified (becoming steeper) 
due to the rapidity dependence of cluster yields (see Fig.~\ref{dndy_dhel}),
thus making a reliable estimate of $\langle m_t\rangle -m$ and its uncertainty impossible.
\begin{figure}
\includegraphics[width=0.85\linewidth]{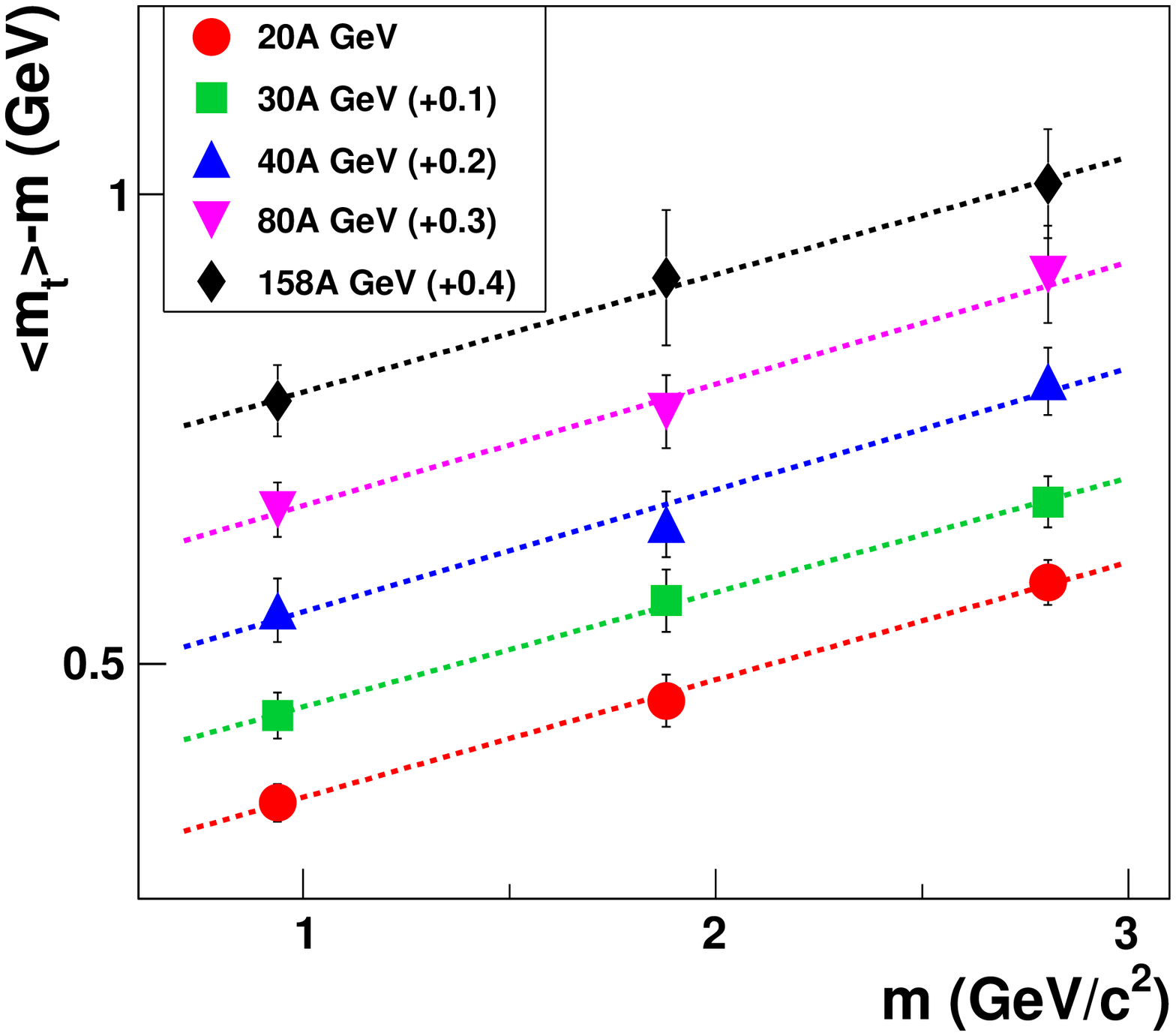}
\caption{(Color online) Mass dependence of $\langle m_t\rangle -m$
in central Pb+Pb collisions at 20$A$-158\agev.
Linear fits to the data points are indicated by dashed lines.}
\label{meanmt_mass}
\end{figure}

In central heavy-ion collisions the pressure gradient in the
system generates strong transverse radial flow. Particles inside a collective
velocity field acquire additional momentum proportional to the
particle's mass. This implies that the average transverse kinetic energy $\langle E_t\rangle_{kin}$ depends
on both the strength of radial flow and random thermal motion as
\begin{widetext}
\begin{equation}\label{therm}
\langle m_t\rangle-m=\langle E_t\rangle_{kin} \approx \langle E \rangle _{therm.}+\langle E \rangle _{flow}=\frac{3}{2}T+(\gamma-1)m~~,
\end{equation}
\end{widetext}
where $\gamma=1/\sqrt{1-\langle \beta\rangle^2}$ and 
$\langle\beta\rangle$ is the average radial collective velocity and $T$ the temperature.
%\end{widetext}
Figure~\ref{meanmt_mass} shows the NA49 results for the mid-rapidity value of
$\langle m_t\rangle -m$ for protons and light nuclei from central
Pb+Pb collisions at 20$A$-158\agev. The data points
for protons were taken from Refs.~\cite{na49_ppbar,na49_ppbar_milica}, the values
for $d$ and $^3$He were obtained in this study.
Evidently, $\langle m_t\rangle -m$ rises approximately linearly
with mass at all collision energies. These results may look surprising because
it seems unlikely that objects of a few MeV binding energy per nucleon are
participating in multiple thermalization collisions which generate the common
velocity field inside fireballs of about 120-140 MeV temperature.
On other hand, it was demonstrated in Ref.~\cite{polleri1} that in the framework of the coalescence approach
the choice of a suitable parametrization for the spatial dependence of the single nucleon
density can reproduce
the observed mass dependence of the inverse slope parameter $T_{eff}$ (or $\langle m_t\rangle -m$)
of composites.  For example, an interplay between a linear
collective flow profile and a uniform density distribution gives an effective
temperature rising linearly with mass.

\begin{figure}
\includegraphics[width=0.95\linewidth]{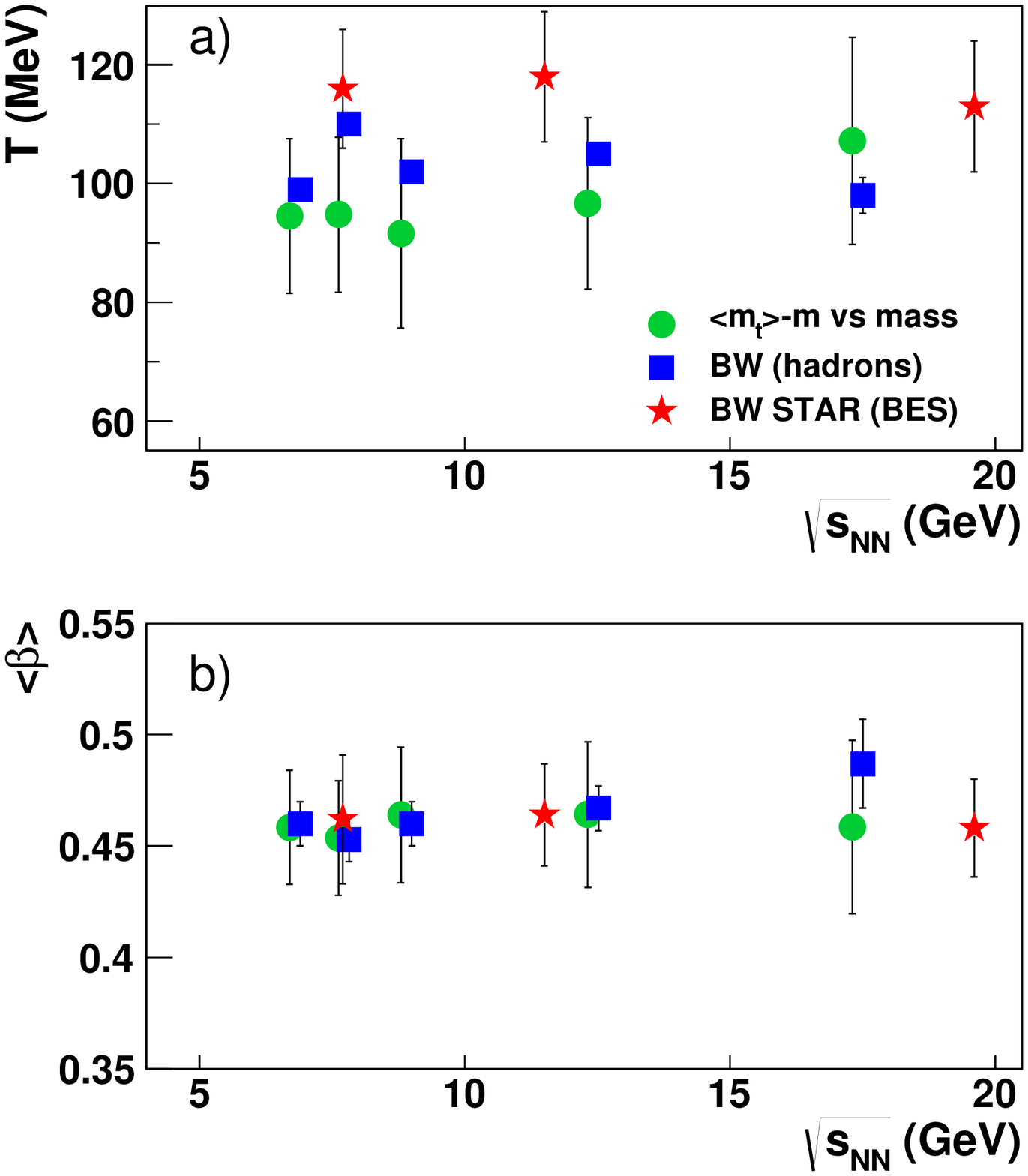}
\caption{(Color online) Energy dependence of the source temperature
$T$ (a) and average collective transverse velocity $\langle \beta \rangle$ (b)
at the kinetic freezeout in central A+A collisions.
The NA49 data from the $m_t$ versus mass analysis (see text for detail)
are indicated by green circles; those from blast-wave (BW) fits of $m_t$ spectra
of hadrons from NA49 are depicted by blue squares; red stars are the STAR-BES results
from a BW analysis of hadron spectra reported in~\cite{star_bw}.}
\label{mt_betat}
\end{figure}

In order to separate the contributions from random thermal and radial
collective motion the data on $\langle m_t \rangle -m$ at each collision
energy  in Fig.~\ref{meanmt_mass} were tested against Eq.~\ref{therm} with
two fit  parameters: $T$ and $\langle \beta \rangle$.
However, as noted in~\cite{heinz_slope}, the extrapolation
of linear fits to zero mass (i.e. the temperature parameter $T$) cannot be directly
related to the source temperature since the apparent temperature in expanding fireballs
is blue shifted as
\begin{equation}\label{form_blast}
T^{*}=T\sqrt{\frac{1+\langle\beta\rangle}{1-\langle\beta\rangle}}
\end{equation}
Thus, in order to obtain the {\it true temperature}, the first fit parameter was
corrected by the blue-shift factor according to Eq.~\ref{form_blast}.
The average transverse velocity $\langle \beta \rangle$ and source temperature
at the kinetic freezeout extracted from these fits are given in Table~\ref{table_slopes}
and plotted in Fig.~\ref{mt_betat} with green circles.

\begin{table}[tb]%[H] add [H] placement to break table across pages
\begin{center}
\begin{ruledtabular}
\caption{\label{table_slopes}Fireball temperature $T$ and mean radial velocity
 $\langle \beta \rangle $ in central Pb+Pb collisions at 20$A$-158\agev~for two different analysis
(see text for detail).}
\begin{tabular}{ccc}
E$_{beam}$ (\agev) & \hspace{1mm}$T$ (MeV) \hspace{1mm}&\hspace{1mm} $\langle \beta \rangle $\hspace{1mm} \\
\hline
\multicolumn{3}{c}{$\langle m_t \rangle -m$ versus mass analysis}\\
20 &\hspace{1mm} $95 \pm 13$ \hspace{1mm} & \hspace{1mm} $0.46 \pm 0.03$ \hspace{1mm}\\
30 &\hspace{1mm} $95 \pm 13$ \hspace{1mm}&\hspace{1mm} $0.45 \pm 0.03$ \hspace{1mm}\\
40 &\hspace{1mm} $92 \pm 15$ \hspace{1mm} &\hspace{1mm} $ 0.46 \pm 0.03$ \hspace{1mm}\\
80 &\hspace{1mm} $97 \pm 14$ \hspace{1mm}&\hspace{1mm} $0.46 \pm 0.03$ \hspace{1mm} \\
158 &\hspace{1mm} $107 \pm 17$ \hspace{1mm}&\hspace{1mm} $0.46 \pm 0.04$ \hspace{1mm}\\
\multicolumn{3}{c}{Blast-Wave (hadrons) analysis}\\
20 &\hspace{1mm} $99 \pm 1$ \hspace{1mm} & \hspace{1mm} $0.46 \pm 0.02$\hspace{1mm}\\
30 &\hspace{1mm} $110 \pm 1$ \hspace{1mm}&\hspace{1mm} $0.45 \pm 0.02$ \hspace{1mm}\\
40 &\hspace{1mm} $102 \pm 1$ \hspace{1mm} &\hspace{1mm} $ 0.47 \pm 0.01$ \hspace{1mm}\\
80 &\hspace{1mm} $105 \pm 1$ \hspace{1mm}&\hspace{1mm} $0.47 \pm 0.01$\hspace{1mm} \\
158 &\hspace{1mm} $98 \pm 2$ \hspace{1mm}&\hspace{1mm} $0.49 \pm 0.02$\hspace{1mm}\\
\end{tabular}
\end{ruledtabular}
\end{center}
\end{table}

\begin{figure}
\includegraphics[width=0.95\linewidth]{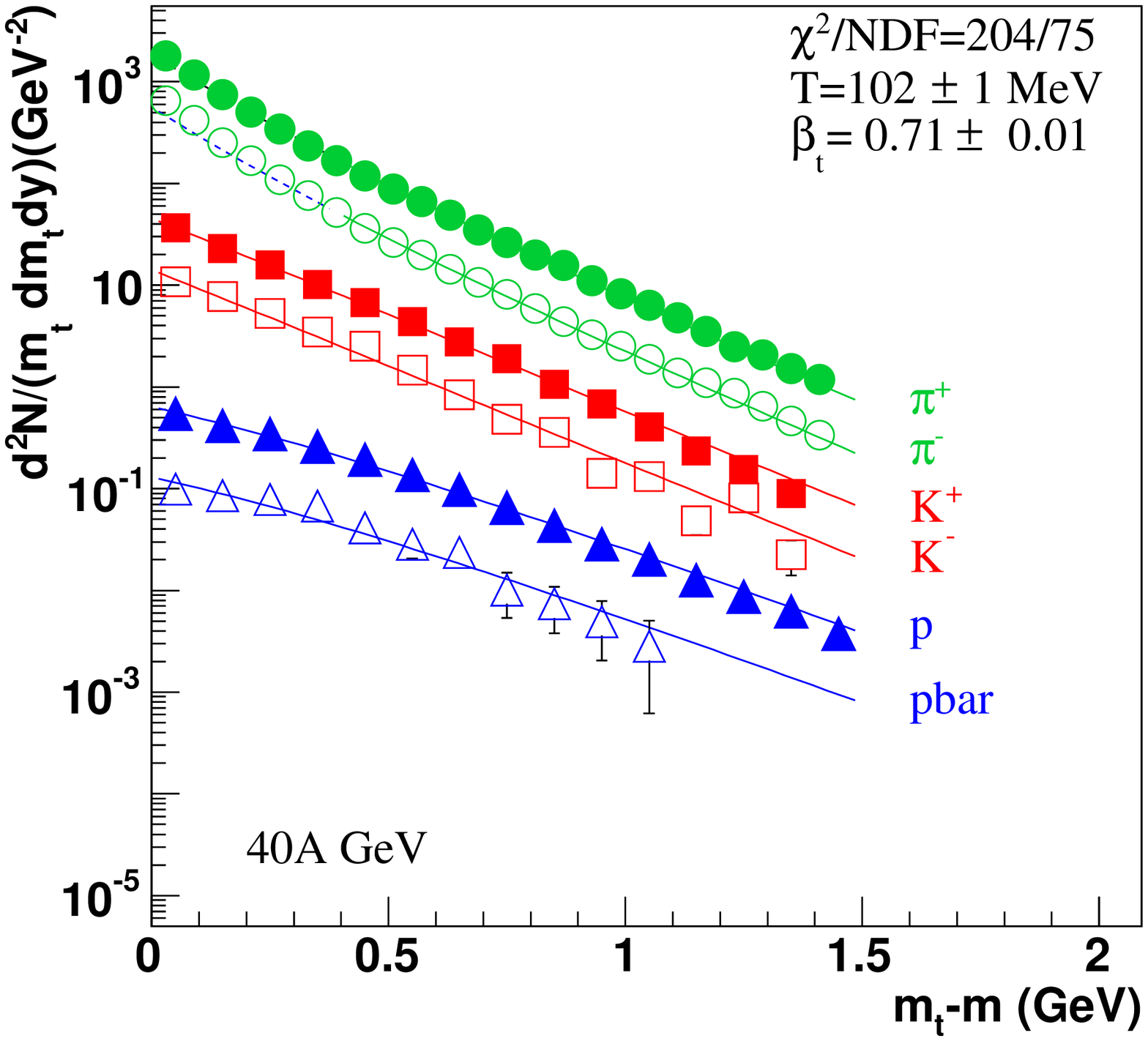}
\caption{(Color online) Blast-Wave (BW) motivated fits to mid-rapidity
$m_t$ spectra of $\pi,K$, $p$ and $\bar{p}$ from central Pb+Pb collisions at 40\agev.
Data from~Refs.~\cite{na49_pika,na49_pika1,na49_ppbar}}
\label{bw_all}
\end{figure}

The discussed source parameters $T$ and $\langle\beta\rangle$ can also be estimated
in the framework of a hydrodynamically inspired blast-wave
(BW) model~\cite{heinz_slope} by fitting the transverse mass spectra of particles
of different masses simultaneously to the function
\begin{widetext} 
\begin{equation}\label{blast}
\frac{d^2N_i}{m_tdm_tdy}=C_i\int_{0}^{1}m_t f(\xi)K_1\left(\frac{m_t \cosh{(\rho)}}
{T}\right) I_0\left(\frac{p_t \sinh{(\rho)}}{T} \right)\xi d\xi~~~,
\end{equation}
\end{widetext}
where $C_i$ is the normalization for particle of type $i$ and $T$ is the freezeout temperature.
The parameter $\rho$ is defined as $\rho=\tanh^{-1}{(\beta_t\xi^n)}$, where $\beta_t$
is the surface velocity and $\xi=r/R$ with $R$ the fireball radius.
Furthermore, a box-like spatial density distribution ($f(\xi)=1$) and a linear
velocity profile ($n=1$) were assumed and thus $\langle \beta \rangle=\frac{2}{3}\beta_t$.
As an example, the results of a BW-analysis of the NA49 experimental data
on charged $\pi$ and $K$ mesons as well as protons and
antiprotons~\cite{na49_pika,na49_pika1,na49_ppbar} from central Pb+Pb collisions
at 40\agev~are shown in Fig.~\ref{bw_all}. BW fits are drawn by
solid curves and fit parameters ($T,\beta_t$) are listed in the inset.
The systematic uncertainties of the fit parameters were estimated
by varying the lower bound of the fitting interval for some species and
by excluding different particles from the analysis. These uncertainties do not
exceed 3-4\% in most cases. The results of the BW fits are tabulated in
Table~\ref{table_slopes} and plotted in Fig.~\ref{mt_betat} by blue squares.
In addition, recent data from the STAR Beam Energy Scan (BES)
program~\cite{star_bw} are shown by red stars. As the results of different
analyses consistently indicate, the freezeout kinetic parameters ($T_{kin},\langle \beta \rangle$)
do not vary significantly within the energy range $6<\sqrt{s_{NN}}<20$~GeV.
 
\subsection{$t/^3$He ratio}
\label{trhel_rat}

Ratios of the yields of nuclear clusters with the same $A$ but different nucleon content
(such as the ratio $t/^3$He) can serve as an indicator of the isospin asymmetry
in the source. The initial $n/p$ ratio of 1.54 in lead nuclei
can vary dramatically in the course of Pb+Pb reactions. During the hadron
phase, multiple nucleon-nucleon and pion-nucleon inelastic collisions inside
the interaction zone change this ratio. The value of $n/p$ at freezeout
can be deduced from comparing the yield of tritons (a composite of two neutrons
and one proton) to that of $^3$He clusters (two protons
and one neutron) because the yield of each species is proportional to different
combinations of the phase space densities of the isospin partners.

\begin{figure}
\includegraphics[width=1.0\linewidth]{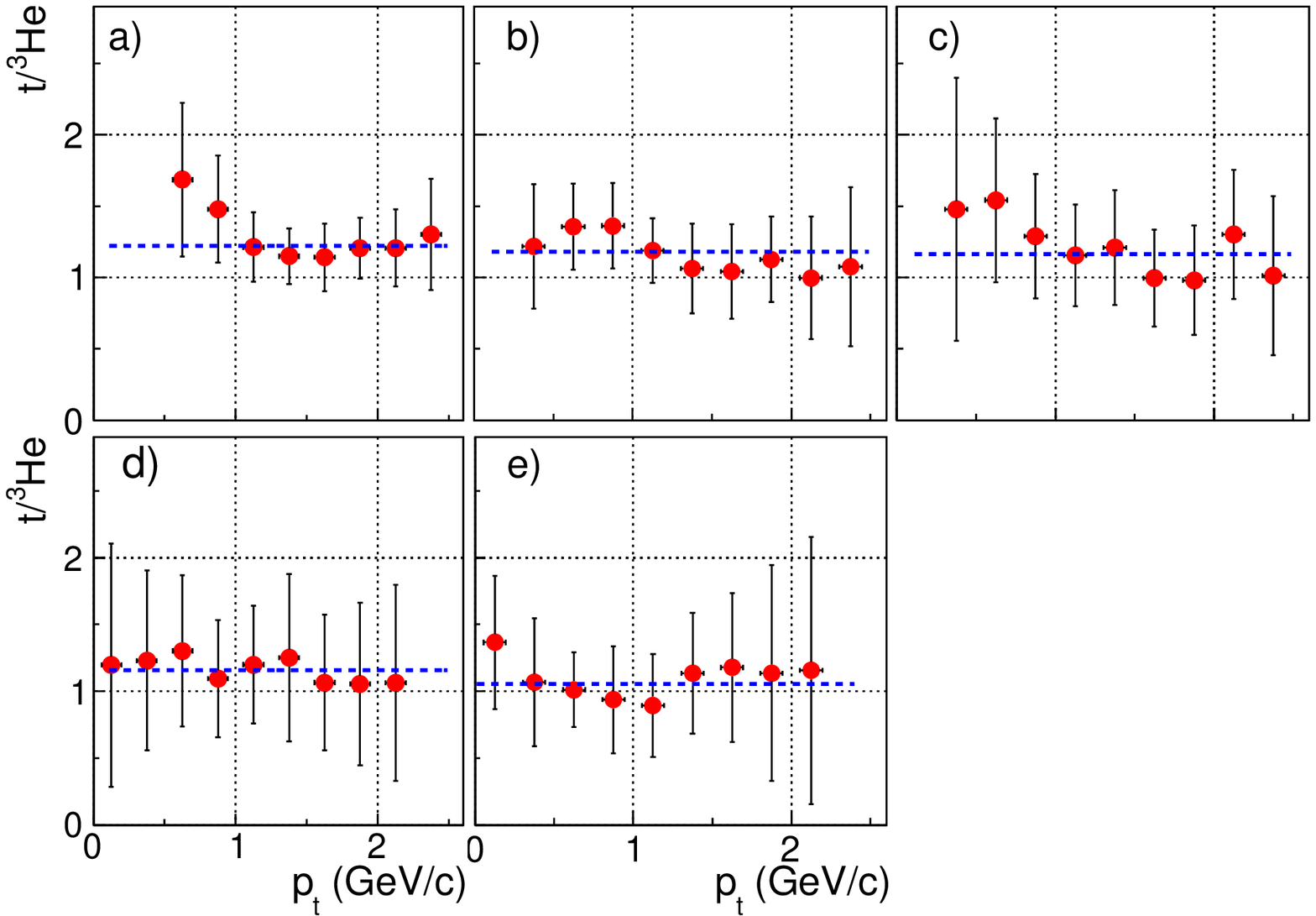}
\caption{(Color online) Ratio of $t$ to $^3$He yields as a function
of $p_t$ from central Pb+Pb collisions at 20$A$ (a), 30$A$ (b), 40$A$ (c),
80$A$ (d), and 158\agev (e).
The dashed lines show the results of fits to a constant.}
\label{trit_hel_rat}
\end{figure} 
 
\begin{figure}
\includegraphics[width=0.8\linewidth]{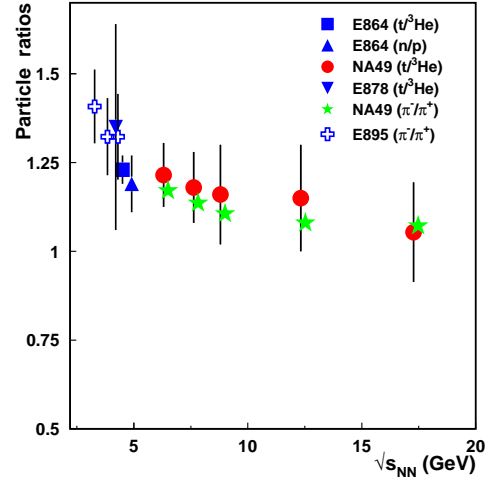}
\caption{(Color online) $n/p$, $t/^3$He, and $\pi^-/\pi^+$ ratios
in central A+A collisions.}
\label{trithel_rat}
\end{figure}

For extracting information on the $n/p$ ratio the shapes of the transverse
momentum distributions for $t$ and $^3$He are studied first. Figure~\ref{trit_hel_rat}
presents the ratio of yields of $t$ to $^3$He as function of $p_t$.
For this particular study, the data for $^3$He at each beam energy
were averaged over the rapidity range of the measurements for tritons.
Because of the TOF acceptance (see Fig.~\ref{acc_all}) the average
ra\-pi\-di\-ty for tritons depends on $p_t$ for small transverse momenta
("banana"-like acceptance) and the dependence is stronger at low beam energies.   
To avoid extra complications due to the change of yields with rapidity,
the ratio was computed above $p_t\approx0.5$~{\gevc} and 0.3~{\gevc}
at 20$A$ and 30$A$-40\agev, respectively. The uncertainties shown in Fig.~\ref{trit_hel_rat}
are mainly associated with the triton statistics and within
these uncertainties there is no evident trend with $p_t$ in the ratio.
For each beam energy the dependence of the $t/^3$He ratio
was fitted to a constant indicated
by dashed lines in Fig.~\ref{trit_hel_rat}. The ratio of triton to $^3$He yields
averaged over the transverse momentum interval 0.3(0.5)$<$$p_t$$<$2.5~{\gevc} 
was found to be $1.22\pm0.10$, $1.18\pm0.11$, $1.16\pm0.15$,
$1.15\pm0.19$, and $1.05\pm0.15$ at 20$A$, 30$A$, 40$A$, 80$A$, and 158\agev~,
respectively. The $t/^3$He ratio is plotted in Fig.~\ref{trithel_rat} (red dots) as a function of the
center-of-mass energy. The decreasing trend with $\sqrt{s_{NN}}$ suggests
that a complete isospin equilibration may eventually be achieved at an energy
above the SPS range. The data points from the
E864~\cite{clust_e864,neut_e864} and E878~\cite{deu_e878} experiments
give an impression of how close the $t/^3$He and $n/p$ ratios are at AGS
energies.

It is also expected that in heavy-ion collisions the $n/p$ ratio and the
$\pi^-$/$\pi^+$ ratio should resemble each other since all these species
are involved in the process of dynamical evolution of the overall
isospin balance. Fi\-gu\-re~\ref{trithel_rat} also shows the NA49 data
on the $\pi^-$/$\pi^+$ ratio at mid-rapidity~\cite{na49_pika,na49_pika1} (green stars)
together with the measurement at lower energies from the E895 experiment~\cite{e895_pions}. 
The measurements indicate that, indeed,  both ratios remain coupled
over the AGS and SPS energy ranges.

\subsection{Coalescence}

\begin{figure}
\includegraphics[width=1.0\linewidth]{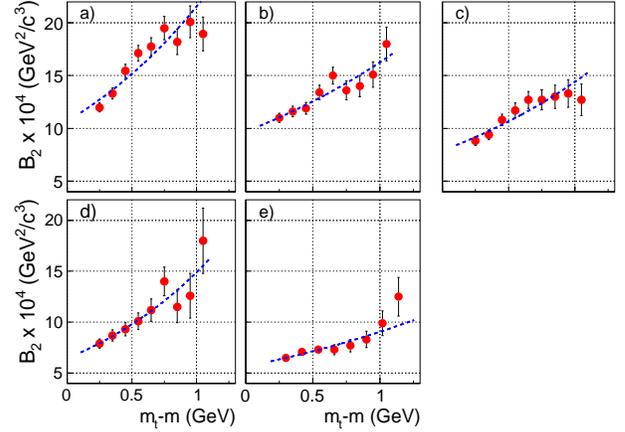}
\caption{(Color online) Coalescence parameter $B_2$ as a function
of $m_t-m$ for deuterons from central
Pb+Pb collisions at 20$A$ (a), 30$A$ (b), 40$A$ (c), 80$A$ (d), and 158\agev (e).
The dashed lines represent fits with an exponential in $m_t$.}
\label{b2_mt}
\end{figure}

\begin{figure}
\includegraphics[width=1.0\linewidth]{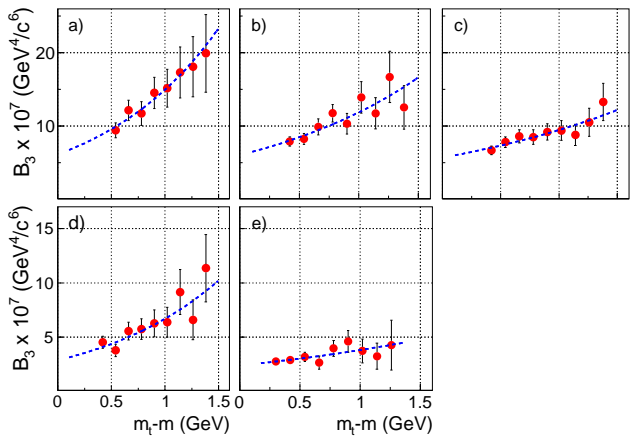}
\caption{(Color online) Coalescence parameter $B_3$ as a function
of $m_t-m$ for $^3$He nuclei from central
Pb+Pb collisions at 20$A$ (a), 30$A$ (b), 40$A$ (c), 80$A$ (d), and 158\agev (e).
The dashed lines represent fits with an exponential in $m_t$.}
\label{b3_mt}
\end{figure}

In a coalescence approach~\cite{coal1,coal2} the invariant yield $N_A$
of clusters with charge $Z$ and atomic mass number $A$
is related to the product of the yields of protons $N_{pr}$ and
neutrons $N_{n}$ through the coefficient $B_A$, the so-called coalescence parameter:
\begin{equation}
\label{eq:b2}
E_A\frac{\der^3N_A}{\der^3P_A}=B_A\left(E_{pr}
\frac{\der^3N_{pr}}{\der^3p}\right)^Z\left(E_{n}
\frac{\der^3N_{n}}{\der^3p}\right)^{A-Z}~~,
\end{equation}
where $p=P_A/A$. Assuming that the ratio of neutrons to protons is unity, $B_A$ is then
calculated by dividing the cluster yield at a given momentum $P_A$ by the
A-th power of the proton yields at $P_A/A$. Results of such a combined
analysis of clusters from this study and the proton spectra measured
in Ref.~\cite{na49_ppbar} are presented in Figs.~\ref{b2_mt}~and~\ref{b3_mt},
which show $B_2$ and $B_3$ as a function of transverse mass at five
collision energies. 
It should be noted, that in a coalescence analysis the data used
for clusters and protons need to be measured in the same rapidity interval
since there is in general a non-negligible rapidity dependence of the particle yields at a given
$m_t$. The available NA49 spectrometer acceptance, however, allows a common $m_t$
coverage only in the region of cluster $m_t$$-m$~$>$~0.25~GeV.
    
It is seen that for all collision energies the coalescence parameters are rising
with transverse mass in accordance with the expectation that strong position-momentum
correlations are present in the expanding source leading to a higher coalescence probability
at larger values of $m_t$~\cite{heinz}.

When calculating the systematic uncertainty of the presented
values of $B_{2,3}$, an uncertainty associated with the proton yields needs to be included. This
was estimated by comparing the NA49 results on proton yields obtained with two different
analysis me\-thods. The \dndy~values for protons from an analysis using \dedx~measurements
reported in Ref.~\cite{na49_ppbar_milica} differ from those based on the combined \dedx+TOF analysis
published in Ref.~\cite{na49_ppbar} by 5\% and 6\% at 40\agev~and 158\agev, respectively.
Based on these differences a systematic uncertainty of 6\% was assigned to the proton yields 
and was further assumed not to vary with energy. Standard error propagation then led 
to an estimated uncertainty of 12\% and 18\% for B$_2$ and B$_3$, respectively.

\begin{figure}
\includegraphics[width=0.9\linewidth]{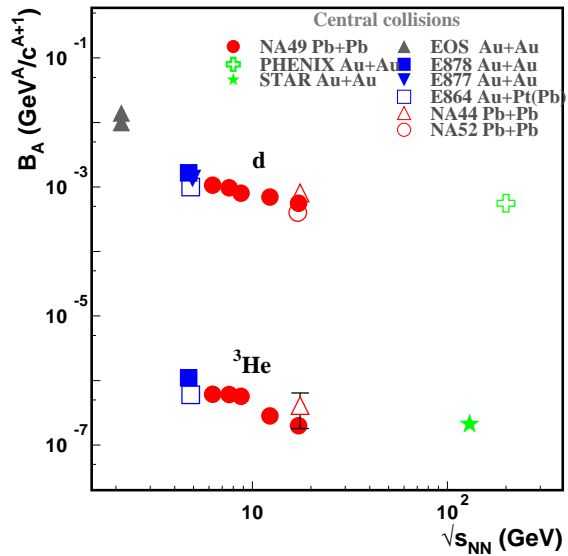}
\caption{(Color online) Coalescence parameter $B_2$ and $B_3$
from central A+A collisions.}
\label{b3_energy}
\end{figure}

Published~results on coalescence factors in heavy-ion experiments have been measured
in different phase space regions since the experiments differed in their rapidity and $p_t$
coverage. In order to compare the present measurements for $B_{2}$ and $B_3$ with previously
obtained results, the dependences on $m_t-m$ shown in Figs.~\ref{b2_mt}~and~\ref{b3_mt} 
were extrapolated down to $m_t-m = 0$ ($p_t=0$).
For this purpose, a functional form of $a_0\exp[a_1(m_t-m)]$
was fitted to the results obtained at each energy and is plotted
by dashed lines. The fit parameter $a_0$ equals the coalescence parameter at $p_t = 0$, while the value of
the parameter $a_1$ depends on the difference between the slope parameters of the
spectra for clusters and protons (i.e. $a_1\approx(1/T_{prot}-1/T_A)$). The results on $B_A$
at $p_t=0$ are listed in Table~\ref{table_coal} and plotted in Fig.~\ref{b3_energy}. 
  
As was pointed out in the introduction, the lack of knowledge about the
production of neutrons in heavy-ion reactions introduces a bias in the determination of the
coalescence parameters when employing only the proton yield.
Using the results on the $t$/$^3$He ratio from Section~\ref{trhel_rat} one can correct
the values for $B_{2,3}$ obtained under the assumption of equal yields for nucleons of both types.
The results in parentheses in Table~\ref{table_coal} are the 
coalescence parameters for $d$ and $^3$He corrected by the ratio $R_{np} \approx t/^3{\text{He}}$.
         
\begin{table}[tbh]%[H] add [H] placement to break table across pages
\begin{center}
\caption{\label{table_coal}Coalescence parameters $B_{2,3}$ at $p_t=0$ from central
Pb+Pb collisions at beam momenta 20$A$-158A~\gevc. The numbers in parentheses are
corrected for the $n/p$ ratio by the factor $R_{np} \approx t/^3{\text{He}}$ (see text).}
\begin{ruledtabular}
\begin{tabular}{ccc}
E$_{beam}$ & $B_2\cdot 10^{4}$ & $B_3\cdot 10^{7}$ \\
(\agev) & (GeV$^2$/c$^3$) &(GeV$^4$/c$^6$)\\
\hline
20 & ~~~$10.7(8.8)\pm0.4$~~~  & ~~~$6.1(5.0)\pm 1.1$~~~ \\
30 & ~~~$9.7(8.2)\pm 0.5$~~~  & ~~~$6.1(5.2)\pm 0.7$~~~ \\
40 & ~~~$7.9(6.8)\pm 0.4$~~~  & ~~~$5.7(4.9)\pm 0.7$~~~ \\
80 & ~~~$6.4(5.6)\pm 0.5$~~~  & ~~~$2.8(2.4)\pm 0.3$~~~\\
158 &~~~$5.6(5.3)\pm 0.4$~~~  &~~~$2.0(1.9)\pm 0.4$~~~ \\
\end{tabular}
\end{ruledtabular}
\end{center}
\end{table}

Within the SPS energy range the variation of the coalescence parameter
is less than 40\% and 60\% for $B_2$ and $B_3$, respectively.
Figure~\ref{b3_energy} compares the results for $B_2$ and $B_3$
at $p_t=0$ (not corrected by $R_{np}$) obtained here to experimental data from the
Bevalac~\cite{clust_e802}, AGS~\cite{clust_e864,clust_e877}, SPS~\cite{deu_na52,deu_na44},
and RHIC~\cite{dbar_star,ddbar_phenix,ddbar_brahms}.
One concludes from this compilation that the coalescence parameters decrease only slowly
with $\sqrt{s_{NN}}$ over a broad range of collision energies.
 
In the framework of thermal models of cluster production~\cite{capusta,mekyan}
the coalescence parameter is a measure of the source size: $B_A\approx (1/V)^{A-1}$.
Thus, the observed ener\-gy dependence of $B_A$ implies that the transverse size
of the emitting source does not change much in this energy domain. 
This behavior is consistent with that found in two-pion interferometry (HBT) 
measurements~\cite{na49_hbt}.
 
\begin{figure}
\includegraphics[width=0.9\linewidth]{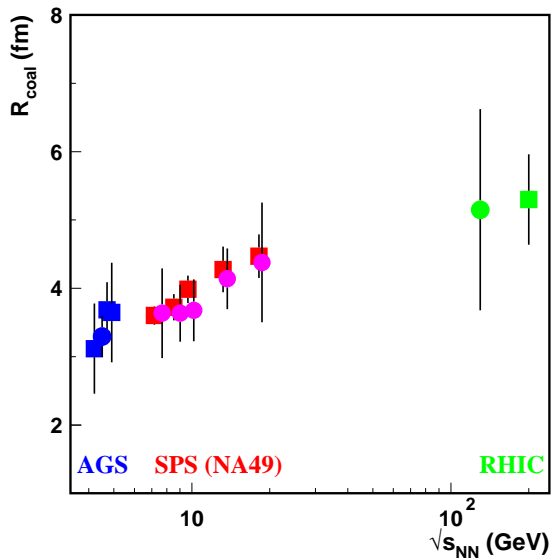}
\caption{(Color online) Coalescence radii $R_{coal}$ for A=2 (squares)
and A=3 (circles) nuclei from central A+A collisions.}
\label{rad_coal}
\end{figure}

In Ref.~\cite{heinz}, calculations implementing collective expansion
of the reaction zone within the density matrix formalism demonstrated
a close relation of the HBT radii to those obtained from the coalescence analysis.
Using the prescription given in Ref.~\cite{heinz} the coalescence radii
($R_{coal}$) for deuterons and $^3$He were calculated at all collisions ener\-gies.
The results are shown in Fig.~\ref{rad_coal} along with the data from the AGS and RHIC. 
One observes that the values of $R_{coal}$
for $d$ and $^3$He agree within their uncertainties and increase gradually with the
collision energy. The latter may indicate a small increase of the freezeout
volume in this energy domain. 

\section{Summary}
This paper presents results on the production of $d$, $t$ and $^3$He nuclei
in central Pb+Pb collisions at 20$A$-158\agev~recorded with the NA49 detector at the CERN SPS.
The results for $^3$He cover a wide range of rapidity and transverse momentum, while 
the measurements for $d$ and $t$ were possible only in regions closer to  mid-rapidity
and more restricted in transverse momentum.
Cluster yields were determined and exhibit a concave shape as function of rapidity
with an increase of the degree of concavity for heavier systems. The
yields of  $d$ and $^3$He integrated over the full phase space agree
with thermal model predictions at all collision energies.
The transverse mass spectra of clusters were measured and
the average values $<$$m_t$$>$$-m$ were found to increase linearly with
the mass. This behavior favors a combination of a box density
profile with a linear velocity profile in the source of the clusters.
The evolution of the isospin asymmetry in the fireball was studied
using the triton to $^3$He  ratio. It was found to change gradually with
collision energy following the trend observed in the ratio of $\pi^-$ to $\pi^+$ yields.     
The coalescence parameters $B_{2,3}$ were derived showing a weak energy dependence.
This observation suggests only a small increase of the freezeout volume from AGS to RHIC
energies.

\begin{acknowledgments}
This work was supported by
the US Department of Energy Grant DE-FG03-97ER41020/A000,
the Bundesministerium fur Bildung und Forschung, Germany (06F~137),
the Virtual Institute VI-146 of Helmholtz Gemeinschaft, Germany,
the Polish Ministry of Science and Higher Education (1~P03B~006~30, 1~P03B~127~30, 
0297/B/H03/2007/33, N~N202~078738), National Science Centre,
Poland (Grant No. 2014/14/E/ST2/00018), the Hungarian Scientific Research Foundation
(T032648, T032293, T043514),
the Hungarian National Science Foundation, OTKA, (F034707),
the Bulgarian National Science Fund (Ph-09/05),
the Croatian Ministry of Science, Education and Sport (Project 098-0982887-2878)
and Stichting FOM, the Netherlands.
\end{acknowledgments}

%==============================================

\end{document}